\documentclass[preprint,11pt]{elsarticle}

\usepackage{amssymb}
\usepackage{amsthm}
\usepackage{mathtools}

\usepackage{color}
\usepackage{amsmath}
\usepackage{comment}

\newcommand{\bx}{\mathbf{x}}
\newcommand{\bv}{\mathbf{v}}
\newcommand{\bB}{\mathbf{B}}
\newcommand{\bb}{\mathbf{b}}
\newcommand{\bE}{\mathbf{E}}

\newcommand{\bu}{\mathbf{u}}
\newcommand{\bj}{\mathbf{j}}

\newcommand{\bh}{\mathbf{h}}
\newcommand{\Dt}{\Delta t}

\newcommand{\deriv}[2]{\frac{d #1}{d #2}}

\newcommand{\order}[1]{\mathcal{O} \left( #1 \right)}

\newcommand{\rd}{\,\mathrm{d}}

\DeclareMathOperator*{\argmin}{arg\,min}

\journal{Journal of Computational Physics}

\begin{document}

\begin{frontmatter}



\title{An explicit, energy-conserving particle-in-cell scheme for relativistic plasmas}


\author[LLNL]{Lee F.\ Ricketson}

\affiliation[LLNL]{organization={Lawrence Livermore National Laboratory, Center for Applied Scientific Computing},
            addressline={7000 East Avenue}, 
            city={Livermore},
            postcode={94550}, 
            state={CA},
            country={USA}}

\author[UW]{Jingwei Hu}

\affiliation[UW]{organization={Department of Applied Mathematics, University of Washington},
            addressline={Box 353925}, 
            city={Seattle},
            postcode={98195}, 
            state={WA},
            country={USA}}

\begin{abstract}
We extend the recently-developed explicit, energy-conserving particle-in-cell (PIC) scheme of \cite{ricketson2025explicit} to the relativistic Vlasov-Maxwell system.  As in the non-relativistic case, the method is built on an optimization problem that is analytically solvable, local to each particle, and designed to enforce exact energy conservation.  Although the solution to this optimization problem is not guaranteed to be real, we show that such instances are rare enough for practical simulation parameters to permit dramatic improvements in energy conservation over traditional explicit PIC schemes.  We show that, as in the non-relativistic case, the scheme is compatible with popular field-solvers for electromagnetic PIC schemes, including the Yee/FDTD and pseudo-spectral analytic time-domain (PSATD) methods.  The scheme is verified on standard relativistic test problems, where its conservation properties are confirmed.  
\end{abstract}




\end{frontmatter}


\section{Introduction}
\label{sec:intro}

Particle-in-cell (PIC) schemes are widely used to simulate kinetic plasma phenomena in a variety of scenarios.  Such schemes are simple, scalable, and relatively robust.  However, their speed and accuracy have historically been denigrated by a lack of discrete conservation properties, most famously leading to so-called ``grid heating" arising from the finite grid instability \cite{barnes2021finite, birdsall2018plasma}.  

There has been considerable recent work on PIC schemes that conserve total energy exactly, thereby eliminating grid heating (and effectively mitigating the finite grid instability itself in many practical contexts \cite{barnes2021finite}).  These schemes have largely been fully implicit \cite{chen2011energy, chacon2016curvilinear, chen2015multi} or semi-implicit \cite{chen2020semi, lapenta2017exactly, markidis2011energy, bacchini2019relativistic}.  More recently, a few efforts have captured exact energy conservation in fully explicit schemes \cite{gonoskov2024explicit, ji2023asymptotic}, including in the authors' precursor to this work \cite{ricketson2025explicit}.  

Much of this development has naturally begun in the non-relativistic limit, where the governing PDEs and resulting dynamics are somewhat simplified.  However, relativistic effects play an important role in several applications of physical interest, including runaway electrons in tokamaks \cite{breizman2019physics}, inertial confinement fusion \cite{atzeni2005fluid}, and astrophysical plasmas \cite{nishikawa2021pic}.  This has motivated development of semi-implicit \cite{chen2020semi} and explicit \cite{gonoskov2024explicit} energy-conserving PIC schemes for the relativistic Vlasov-Maxwell system.  

Here, we extend the scheme of \cite{ricketson2025explicit} to the relativistic case.  Compared to the semi-implicit scheme of \cite{chen2020semi}, a reduced cost per time-step is expected at the expense of the ability to step over stiff time-scales in the particle advance. Although both \cite{gonoskov2024explicit} and the present scheme are fully explicit, \cite{gonoskov2024explicit} relies on a splitting scheme that requires that particles within an individual cell be advanced in serial. The present scheme has no such requirement, and thus improved parallel scalability of the present scheme is expected in some contexts.  

As in \cite{ricketson2025explicit}, the scheme is built on a standard time integration scheme with an additional correction at the end of each time-step that enforces energy conservation.  The correction is defined as the solution to a constrained optimization problem, with the objective function preserving accuracy by minimizing distance to a velocity with known accuracy, and the constraint enforcing exact conservation.  Importantly, this optimization problem is analytically solvable, thus preserving the explicit nature of the scheme, and local to each particle, thus preserving parallel scalability.  

As in the non-relativistic case, we show that the new time-integration scheme is compatible with popular Maxwell solvers from the relativistic PIC literature.  In particular, energy conservation can be achieved with both the Yee grid/finite difference time domain (FDTD) scheme \cite{Yee66} and the pseudospectral analytic time domain (PSATD) scheme \cite{birdsall2018plasma, lehe2018review, vay2013domain}.  Each scheme is desirable for its accurate (and in the case of PSATD, exact) capturing of the light-wave dispersion relation, which is of particular importance in relativistic applications.  

The scheme is verified on four test problems that feature relativistic effects: the relativistic two-stream instability, relativistic Landau damping, the filamentation instability, and the relativistic Weibel instability.  In each case, agreement with standard PIC schemes and/or linear theory is observed, along with dramatically improved energy conservation.  

The remainder of the article is structured as follows.  In Section \ref{sec:background}, we review relevant background material, including the relativistic Vlasov-Maxwell system, its standard PIC discretizations, and the previous energy-conserving scheme in the non-relativistic case.  In Section \ref{sec:method}, we derive the scheme, showing both that it retains second-order accuracy and reduces to the previous scheme in the non-relativistic limit.  Numerical results confirming the theoretical predictions are shown in Section \ref{sec:numerics}, and we conclude in Section \ref{sec:conclusions}.  


\section{Background}
\label{sec:background}

\subsection{Relativistic Vlasov-Maxwell system}

This work concerns numerical solution of the relativistic Vlasov-Maxwell system.  The relativistic Vlasov equation is given by
\begin{equation}
\label{vlasov}
    \partial_t f + \bv \cdot \nabla_\bx f + \left( \bE + \bv \times \bB \right) \cdot \nabla_\bu f = 0.
\end{equation}
Here, $f(\bx, \bu, t)$ is the phase space particle distribution function.  $\bx$ denotes position in physical space, $\bv$ denotes particle velocity, and $\bu = \bv \gamma$ denotes proper velocity, with the Lorentz factor 
\begin{equation}
    \gamma = \sqrt{1 + \left\| \bu \right\|^2/c^2} = \frac{1}{\sqrt{1 - \left\| \bv \right\|^2/c^2}}.  
\end{equation}
We work here and in the remainder of the paper in a non-dimensional formulation in which time is scaled by the plasma frequency $\omega_p = \sqrt{n_0 e^2/m \epsilon_0}$, space by an arbitrary factor $L$, and velocity by $L \omega_p$.  Here, $n_0$ is a reference number density, $m$ the particle mass, $e$ the fundamental unit charge, and $\epsilon_0$ the permittivity of free space.  Common choices for $L$ include the DeBye length $\lambda_D$ and the distance traveled by light in a plasma period $c/\omega_p$, but we need not specify the choice here.  The resulting normalization factors for $f$, $\bE$, and $\bB$ are, respectively, $n_0/(\omega_p L)^3$,  $n_0Le/\epsilon_0$ and $n_0e/\epsilon_0 \omega_p$.

$\bE$ and $\bB$ denote the electric and magnetic fields, which are specified by Maxwell's equations:
\begin{equation}
\begin{split}
    \nabla_x \cdot \bE &= \rho - 1, \\
    \nabla_x \cdot \bB &=0, \\
    \nabla_x \times \bE &= -\frac{\partial \bB}{\partial t}, \\
    \nabla_x \times \bB &= \frac{1}{c^2} \left( \frac{\partial \bE}{\partial t} + \bj \right).
\end{split}
\end{equation}
These equations couple back to Vlasov via the charge density $\rho$ and current density $\bj$, given by
\begin{equation}
    \rho = \int f \, d\bu, \qquad \bj = \int \bv f \, d\bu.
\end{equation}
The speed of light $c$ here is understood to be written in our dimensionless variables -- that is, the physical speed of light divided by $L \omega_p$.  Note also the inclusion of a neutralizing background in Gauss' law, representing an ion density that is constant on time-scales of interest.  All methods developed here can be trivially extended to the multi-species case.  

This system features a conserved total energy. To see it, we first note that equation \eqref{vlasov}
 can be written in a conservative form:
\begin{equation}
    \partial_t f + \nabla_{\bx}  \cdot  (\bv f) + \nabla_\bu \cdot \left( (\bE + \bv \times \bB)  f \right) = 0.
\end{equation}
Taking the moments $\int \cdot \,(1,\gamma)^T \,d\bu$ of the above equation and using integration by parts, one obtains the local conservation of charge and energy:
\begin{align}
&\partial_t \rho +\nabla_{\bx}\cdot \bj=0, \label{eq:charge}\\
&\partial_t \int \gamma f\,d{\bu}+\nabla_{\bx}\cdot \int \bu f\,\rd{\bu}=\frac{1}{c^2}\bE\cdot \bj.\label{eq:localenergy}
\end{align}
Further integration of \eqref{eq:localenergy} in $\bx$ and assuming periodic or zero boundary condition gives 
\begin{equation}
\partial_t \iint \gamma f\,d{\bu}\,d{\bx}=\int\frac{1}{c^2}\bE\cdot \bj\,d{\bx},
\end{equation}
which, combined with the Maxwell's equations, yields
\begin{equation}
\partial_t \left( c^2\iint \gamma f\,d{\bu}\,d{\bx}+\frac{1}{2}\int \left(\|\bE\|^2+c^2\|\bB\|^2\right)
\,d\bx\right)=0.
\end{equation}
Using that $c^2\iint f\,d\bu d\bx$ is conserved (from integration of \eqref{eq:charge} in $\bx$), we can see that the total energy defined by
\begin{equation} \label{eq:relenergy}
    \mathcal{E} = c^2 \int (\gamma - 1) f \, d\bu d\bx + \frac{1}{2} \int \left( \left \| \bE \right\|^2 + c^2 \left\| \bB  \right\|^2 \right) \, d\bx
\end{equation}
is conserved over time. In the non-relativistic limit, $\frac{\bu}{c}\rightarrow 0$ (or $\frac{\bu}{c}\ll 1$), then $\gamma\rightarrow 1$, and $\bu\rightarrow \bv$. Since $c^2(\gamma-1)=\frac{|\bu|^2}{\gamma+1}$, then $c^2(\gamma-1)\rightarrow \frac{|\bv|^2}{2}$ and the above energy reduces to the classical non-relativistic energy.  We seek a discretization that exactly preserves the conservation of \eqref{eq:relenergy}.  

\subsection{Particle-in-cell discretization}

Motivated by the high-dimensionality of the Vlasov equation, particle-in-cell (PIC) schemes work from the ansatz that $f$ can be approximated by a weighted sum of Dirac delta functions:
\begin{equation}
    f(\bx, \bu, t) = \sum_{p=1}^{N_p} w_p \delta \left( \bx - \bx_p(t) \right) \delta \left( \bu - \bu_p(t) \right).
\end{equation}
The evolution equations for the ``particle" states $(\bx_p, \bv_p, w_p)$ are of course the characteristic equations for Vlasov:
\begin{equation}
    \deriv{\bx_p}{t} = \bv_p, \qquad \deriv{\bu_p}{t} = \bE( \bx_p, t ) + \bv_p \times \bB(\bx_p,t), \qquad \deriv{w_p}{t} = 0.
\end{equation}
The electromagnetic fields are computed on a configuration-space mesh with grid-points denoted by $\bx_h$.  Particle data is deposited on the mesh via so-called ``shape functions".  In particular, charge and current densities at grid points are defined by
\begin{equation}
    \rho_h(t) = \frac{1}{| \bh |}\sum_p w_p S^h_\rho \left( \bx_h - \bx_p(t) \right), \qquad \bj_h(t) = \frac{1}{| \bh |}\sum_p w_p \bv_p(t) S^h_j \left( \bx_h - \bx_p(t) \right).
\end{equation}
Here $| \bh |$ denotes cell volume and $S^h$ is a shape function such that $S^h/|\bh|$ is a second-order approximation of the Dirac delta function.  The most commonly used example is the ``tent" function, $S^h(z) = \max \{ 0, 1 - |z|/h \}$ and its tensor products in higher dimensions.  However, arbitrary-order $B$-splines may be used as well.  

$\rho_h(t)$ and $\bj_h(t)$ are used to compute $\bE_h(t)$ and $\bB_h(t)$ via some discretization of Maxwell's equations.  The electromagnetic fields are then interpolated to particle locations:
\begin{equation}
\begin{split}
    \bE(\bx_p, t) &= \sum_h \bE_h(t) S^h_E \left( \bx_p - \bx_h \right), \\
    \bB(\bx_p, t) &= \sum_h \bB_h(t) S^h_B \left( \bx_p - \bx_h \right).
\end{split}
\end{equation}
Typically, some of the shape functions $S^h_\rho$, $S^h_j$, $S^h_E$, and $S^h_B$ are identical.  Energy conservation proofs, in particular, usually require $S^h_j = S^h_E$.  The scheme is finally completed by choosing a temporal discretization of the resulting system of ordinary differential equations.  

It will be instructive to review some of the commonly used discretizations of the characteristic equation and of Maxwell's equations.  We do so in the next two subsections.  

\subsection{Particle advance}
In the non-relativistic limit, the Boris scheme \cite{qin2013boris, HL18} is the \textit{de facto} standard for explicit discretization of the characteristic equation.  In the relativistic case, the Lorentz factor introduces additional discretization choices that have resulted in several similar methods in common use.  They can all be written in the form
\begin{equation} \label{eq:gen_integrator}
\begin{split}
    \bx_p^* &= \bx_p^n + \frac{\Delta t}{2} \bv_p^n, \\
    \bu_p^{n+1} &= \bu_p^n + \Delta t \left( \bE(\bx_p^*, t^{n+1/2} ) + \bar{\bv}_p \times \bB(\bx_p^*, t^{n+1/2}) \right), \\
    \bx_p^{n+1} &= \bx_p^* + \frac{\Delta t}{2} \bv_p^{n+1}.
\end{split}
\end{equation}
Here, $n$ indexes time-step,  $\bv_p^n = \bu_p^n/\gamma_p^n$, $\gamma_p^n = \sqrt{1 + \left\| \bu_p^n \right\|^2/c^2}$ and similar for $\bv_p^{n+1}$.  Three frequently-used schemes differ only in their definition of $\bar{\bv}_p$.  

The relativistic version of the Boris scheme \cite{boris1970relativistic} uses $\bar{\bv}_p = \bu_p^{n+1/2}/\bar{\gamma}_p^{Boris}$, where $\bu_p^{n+1/2} = (\bu_p^n + \bu_p^{n+1})/2$ and 
\begin{equation} \label{eq:relBoris_vbardef}
    \bar{\gamma}_p^{\text{Boris}} = \sqrt{1 + \frac{ \left\| \bu_p^n + \frac{\Delta t}{2} \bE(\bx_p^n, t^n) \right\|^2 }{c^2}}.
\end{equation}
Note that this value of $\bar{\gamma}_p$ is explicitly computable.  Thus, the velocity update in \eqref{eq:gen_integrator} is linearly implicit in the same sense as the non-relativistic version of Boris.  It can thus be analytically inverted using the same techniques.  

However, in the relativistic case, the Boris method does not accurately capture the $\bE \times \bB$ drift.  Vay \cite{vay2008simulation} introduced a modification that does, choosing instead
\begin{equation} \label{eq:Vay_vbardef}
    \bar{\bv}_p^{\text{Vay}} = \frac{1}{2} \left( \frac{\bu_p^{n+1}}{\gamma_p^{n+1}} + \frac{\bu_p^n}{\gamma_p^n}\right).
\end{equation}
Note that this scheme is now \textit{nonlinearly} implicit in the velocity update, since $\gamma_p^{n+1}$ now depends on the updated velocity.  Nevertheless, Vay showed that this relation can be inverted analytically and thus remains effectively explicit in the same sense as Boris.  

While the Vay method corrects the $\bE \times \bB$ drift velocity, it breaks the conservation of phase-space volume enjoyed by the Boris method.  Higuera and Cary \cite{higuera2017structure} introduced a scheme that both conserves phase-space volume and captures the $\bE \times \bB$ drift.  That scheme makes the choice 
\begin{equation} \label{eq:HC_vbardef}
    \bar{\bv}_p^{\text{HC}} = \frac{\bu_p^{n+1/2}}{\gamma_p^{n+1/2}}, \qquad \gamma_p^{n+1/2} = \sqrt{ 1 + \frac{\left\| \bu_p^{n+1/2} \right\|^2}{c^2}}
\end{equation}
Like the Vay scheme, this results in a nonlinearly implicit velocity update.  Also like Vay, Higuera and Cary show that this relation can be inverted analytically using an analogous algebraic process, preserving the effective explicitness, and thus low per-step cost, of the method.  

A review and numerical comparison of all three schemes may be found in \cite{bacchini2019relativistic}.  A review of these and many other methods appears in \cite{schmitz2026overview}, in which it is concluded that while no scheme is universally optimial, Higuera-Cary performs quite well overall.  Also notable for our purposes is the use of a Higuera-Cary-like choice in \cite{chen2020semi} to find a semi-implicit PIC scheme with exact energy conservation.  For both its structure-preserving properties and its convenience in our energy-conservation derivation, we make the Higuera-Cary choice of $\bar{\bv}_p$ in the remainder of this work.  

\subsection{Maxwell discretizations}
\label{sec:maxwell}

It is widely acknowledged that it is important in a variety of contexts to discretize Maxwell's equations in a manner that preserves the analytic light-wave dispersion to the extent possible.  Doing so mitigates numerical Cherenkov radiation, helps capture dephasing in laser-wakefield acceleration \cite{cowan2013generalized}, and captures Doppler harmonics in high-density plasmas subjected to petawatt class lasers \cite{blaclard2017pseudospectral}.  

Two particularly popular methods for doing this are the (a) finite difference time domain (FDTD) and (b) pseudo-spectral analytic time domain (PSATD) schemes.  The former uses a finite difference spatial discretization on Yee's lattice \cite{Yee66, gedney2011yee} in concert with a leapfrog-type temporal discretization.  This leads to light-wave dispersion errors that are tolerable in many scenarios \cite{petropoulos1994phase, schneider2001dispersion}.  The latter uses a pseudospectral spatial discretization, leveraging this description to time-advance the vacuum portion of Maxwell's equations \textit{analytically}, with the only approximation coming from the assumption that the current is constant within a time-step \cite{birdsall2018plasma, vay2013domain}.  This scheme thus captures light wave dispersion \textit{exactly}.  In particular, the Fourier transform of the electromagnetic fields are, given some current $\bj$ considered fixed within a time-step,
\begin{equation} \label{eq:PSATD_update}
\begin{split}
    \mathcal{F} \left[ \bE^{n+1}_h \right] &= C \mathcal{F} \left[ \bE^n_h \right] + i S c \widehat{\mathbf{k}} \times \mathcal{F}\left[ \bB^n_h \right] - \frac{S}{kc} \mathcal{F} \left[ \bj \right]\\
    & \qquad  + (1 - C) \widehat{\mathbf{k}} \left( \widehat{\mathbf{k}} \cdot \mathcal{F}\left[ \bE^n_h \right] \right) + \widehat{\mathbf{k}} \left( \widehat{\mathbf{k}} \cdot \mathcal{F}\left[ \bj \right] \right) \left( \frac{S}{kc} - \Delta t \right), \\
    \mathcal{F}\left[ \bB^{n+1}_h \right] &= C \mathcal{F}\left[ \bB^{n}_h \right] - i\frac{S}{c} \widehat{\mathbf{k}} \times \mathcal{F}\left[ \bE^{n}_h \right] + i \frac{1-C}{kc^2} \widehat{\mathbf{k}} \times \mathcal{F} \left[ \bj \right],
\end{split}
\end{equation}
where $C = \cos (k c \Dt )$, $S = \sin ( k c \Dt )$, and $\mathcal{F}$ denotes the discrete Fourier transform.  The fields themselves are then recovered via an inverse Fourier transform

Most critically for our purposes, each of these schemes respects integration by parts in the sense that for arbitrary functions $\mathbf{F}_h$ and $\mathbf{G}_h$ defined on the mesh, one has
\begin{equation} \label{eq:discreteintegrationbyparts}
    \sum_h \left( \mathbf{G}_h \cdot \nabla_h \times \mathbf{F}_h - \mathbf{F}_h \cdot \nabla_h \times \mathbf{G}_h \right) = 0,
\end{equation}
where $\nabla_h \times$ denotes the discretized curl operator.  This identity is well-known to hold for the Yee lattice \cite{Yee66} and was proved for pseudospectral discretization in Appendix D of \cite{ricketson2025explicit}.  

\subsection{Non-relativistic energy-conserving scheme}
\label{sec:classical_scheme}
The scheme developed in this manuscript is the relativistic extension of the scheme from \cite{ricketson2025explicit}.  We thus find it useful to summarize the development of that scheme, as it strongly informs the relativistic case. Although the original scheme can be applied to either electrostatic or electromagnetic systems, we focus here only on the electromagnetic case.  Finally, the non-relativistic scheme featured two versions.  We present only ``version 2" here, which was shown in \cite{ricketson2025explicit} to feature improved energy conservation compared to ``version 1" at minimal extra cost.  

Even within ``version 2", there are several variants of the scheme depending on how Maxwell's equations are discretized.  The most immediate and natural takes the form
\begin{equation} \label{eq:EMbasic}
\begin{split}
    \bx^*_p &= \bx^n_p + \frac{\Delta t}{2} \bv^n_p, \\
    \bE_h^* &= \bE_h^n + \frac{\Delta t}{2} \left( c^2 \nabla_h \times \bB_h^n -  \bj_h^{n,*} \right), \\
    \bB_h^* &= \bB_h^n - \frac{\Delta t}{2} \nabla_h \times \bE_h^n, \\
    \bv^*_p &= \bv^n_p + \frac{\Delta t}{2} \left( \bE^{*,*}_p + \bv_p^* \times \bB_p^{*,*} \right), \\
    \bx^{n+1}_p &= \bx^n_p + \Delta t \bv^*_p, \\
	\bE_h^{n+1} &= \bE_h^n + \Delta t \left( c^2 \nabla_h \times \bB_h^{n+1/2} -  \bj_h^{*,n+1/2} \right), \\
    \bB_h^{n+1} &= \bB_h^n - \Delta t \nabla_h \times \bE_h^{n+1/2}, \\
    \bv^{\dagger}_p &= \bv^n_p + \Delta t \left( \bE^{n+1/2}_p + \bv_p^* \times \bB_p^{n+1/2} \right), \\
    \bv^{n+1}_p &= \bv_p^\dagger \sqrt{ 1 + 2\frac{ \left( \bv_p^* - \frac{\bv_p^\dagger + \bv_p^n}{2} \right) \cdot (\bv_p^\dagger - \bv_p^n) }{\| \bv_p^\dagger \|^2}},
\end{split}
\end{equation}
where
\begin{equation} \label{eq:jEB_defs}
\begin{split}
    \bj_h^{n,*} &= \frac{1}{\lvert \mathbf{h} \rvert} \sum_p w_p \bv_p^n S^h \left( \bx_p^{*} - \bx_h \right),\\
    \bE^{*,*}_p &= \sum_h \bE_h^* S^h \left( \bx^*_p - \bx_h \right), \\ 
    \bj_h^{*,n+1/2} &= \frac{1}{\lvert \mathbf{h} \rvert} \sum_p w_p \bv_p^* S^h \left( \bx_p^{n+1/2} - \bx_h \right), \\
\bx_p^{n+1/2}&=\frac{\bx_p^n+\bx_p^{n+1}}{2}, \quad \bE_h^{n+1/2} = \frac{\bE_h^n + \bE_h^{n+1}}{2}, \\ 
    \bE^{n+1/2}_p &= \sum_h \bE_h^{n+1/2} S^h \left( \bx^{n+1/2}_p - \bx_h \right), 
\end{split}
\end{equation}
and definitions of the various evaluations of $\bB$ are directly analogous to those for $\bE$.  

Note that this scheme is explicit in the particle update, but that the time-integrator for Maxwell's equations is Crank-Nicolson, making it linearly implicit in the field variables.  The scheme can be made fully explicit with different Maxwell discretizations, which we detail below, but we begin with the simplest case.  

Deriving energy conservation begins with the observation that 
\begin{equation} \label{eq:particle_to_field_id}
\begin{split}
    \sum_p w_p \bv_p^* \cdot \left( \bv_p^\dagger - \bv_p^n \right) &= \Delta t \sum_p w_p \bv_p^* \cdot \left( \bE_p^{n+1/2} + \bv_p^* \times \bB_p^{n+1/2} \right) \\
    &= \Delta t \sum_p \sum_h w_p \bv_p^* \cdot \bE_h^{n+1/2} S^h\left( \bx_p^{n+1/2} - \bx_h \right) \\
    &= \Delta t |\mathbf{h}| \sum_h \bE_h^{n+1/2} \cdot \bj_h^{*,n+1/2}.
\end{split}
\end{equation}
Analysis of the field update shows that
\begin{equation} \label{eq:field_id}
\begin{split}
    \Delta t \sum_h \bE_h^{n+1/2} \cdot \bj_h^{*,n+1/2} &= -\sum_h\bE_h^{n+1/2} \cdot \left( \bE_h^{n+1} - \bE_h^n - \Delta t c^2 \nabla_h \times \bB_h^{n+1/2} \right) \\
    &= -\frac{1}{2}\sum_h \left( \| \bE_h^{n+1} \|^2 - \| \bE_h^n \|^2 \right) \\
    &\qquad + \Delta t c^2 \sum_h \bE_h^{n+1/2} \cdot \nabla_h \times \bB_h^{n+1/2} \\
    &= -\frac{1}{2}\sum_h \left( \| \bE_h^{n+1} \|^2 - \| \bE_h^n \|^2 \right) \\
    &\qquad + \Delta t c^2 \sum_h \bB_h^{n+1/2} \cdot \nabla_h \times \bE_h^{n+1/2} \\
    &= -\frac{1}{2}\sum_h \left( \| \bE_h^{n+1} \|^2 - \| \bE_h^n \|^2 \right) \\
    &\qquad - c^2 \sum_h \bB_h^{n+1/2} \cdot \left( \bB_h^{n+1} - \bB_h^n \right) \\
    &= -\frac{1}{2}\sum_h \left( \| \bE_h^{n+1} \|^2 + c^2 \| \bB_h^{n+1} \|^2 - \| \bE_h^n \|^2 - c^2 \| \bB_h^n \|^2 \right).
\end{split}
\end{equation}
Critically, this line of reasoning relies on the spatial discretization satisfying the integration by parts identity \eqref{eq:discreteintegrationbyparts}.  

Finally, $\bv_p^{n+1}$ in \eqref{eq:EMbasic} is \textit{defined} to be the solution of the optimization problem
\begin{equation} \label{eq:nonrel_opt_def}
    \bv_p^{n+1} = \argmin_{\bv} \left\| \bv - \bv_p^\dagger \right\| \quad \text{s.t.} \quad \frac{1}{2} \left\| \bv_p^{n+1} \right\|^2 - \frac{1}{2} \left\| \bv_p^{n} \right\|^2 = \bv_p^* \cdot \left( \bv_p^\dagger - \bv_p^n \right).  
\end{equation}
It happens that this optimization problem has an analytic solution, given by the expression in the last line of \eqref{eq:EMbasic}.  Thus, one trivially has
\begin{equation} \label{eq:particle_id}
    \sum_p \left( \frac{w_p}{2} \left\| \bv_p^{n+1} \right\|^2 - \frac{w_p}{2} \left\| \bv_p^{n} \right\|^2 \right) = \sum_p w_p \bv_p^* \cdot \left( \bv_p^\dagger - \bv_p^n \right).
\end{equation}

Equations \eqref{eq:particle_to_field_id}, \eqref{eq:field_id}, and \eqref{eq:particle_id} combine to straightforwardly imply
\begin{equation}
\begin{split}
    &\sum_p \frac{w_p}{2} \left\| \bv_p^{n+1} \right\|^2 + \frac{|\mathbf{h}|}{2} \sum_h \left( \left\| \bE_h^{n+1} \right\|^2 + c^2 \left\| \bB_h^{n+1} \right\|^2 \right)  \\
   = &\sum_p \frac{w_p}{2} \left\| \bv_p^{n} \right\|^2 + \frac{|\mathbf{h}|}{2} \sum_h \left( \left\| \bE_h^{n} \right\|^2 + c^2 \left\| \bB_h^{n} \right\|^2 \right),
\end{split}
\end{equation}
which is precisely the statement of discrete energy conservation.  

In \cite{ricketson2025explicit}, it is shown that the same line of reasoning may be adapted to show energy conservation when Maxwell's equations are discretized with either the FDTD or PSATD methods described in Section \ref{sec:maxwell}.  In the case of FDTD, the scheme now reads
\begin{equation} \label{eq:EMleapfrog}
\begin{split}
    \bx^*_p &= \bx^n_p + \frac{\Delta t}{2} \bv^n_p, \\
    \bB_h^{n+1/2} &= \bB_h^{n-1/2} - \Delta t \nabla_h \times \bE_h^{n}, \\
    \bE_h^* &= \bE_h^n + \frac{\Delta t}{2} \left( c^2 \nabla_h \times \bB_h^{n+1/2} -  \bj_h^{n,*} \right), \\
    \bv^*_p &= \bv^n_p + \frac{\Delta t}{2} \left( \bE^{*,*}_p + \bv_p^* \times \bB_p^{n+1/2,*} \right), \\
	\bE_h^{n+1} &= \bE_h^n + \Delta t \left( c^2 \nabla_h \times \bB_h^{n+1/2} -  \bj_h^{*,n+1/2} \right), \\
    \bx^{n+1}_p &= \bx^n_p + \Delta t \bv^*_p, \\
    \bv^{\dagger}_p &= \bv^n_p + \Delta t \left( \bE^{n+1/2}_p + \bv_p^* \times \bB_p^{n+1/2} \right), \\
    \bv^{n+1}_p &= \bv_p^\dagger \sqrt{ 1 + 2\frac{ \left( \bv_p^* - \frac{\bv_p^\dagger + \bv_p^n}{2} \right) \cdot (\bv_p^\dagger - \bv_p^n) }{\| \bv_p^\dagger \|^2}}.
\end{split}
\end{equation}
Directly analogous logic to that followed in \eqref{eq:field_id} allows one to show that the total energy 
\begin{equation} \label{eq:nonstd_energy}
    \mathcal{E}^n = \sum_p \frac{w_p}{2} \left\| \bv_p^{n} \right\|^2 + \frac{|\mathbf{h}|}{2} \sum_h \left( \left\| \bE_h^{n} \right\|^2 + c^2 \bB_h^{n-1/2} \cdot \bB_h^{n+1/2} \right)
\end{equation}
is conserved.  The definition of magnetic potential energy is non-standard, but differs from the standard definition only by $\order{\Delta t^2}$, is almost-surely non-negative, and has precedent in its usage in energy-conserving schemes with the FDTD method \cite{chen2020semi}.  

On the other hand, in the case of PSATD, it is necessary to modify
\begin{equation}
    \bE_p^{n+1/2} \rightarrow \left\langle \bE_p \right\rangle_n^{n+1} \coloneq \frac{1}{\Dt} \int_{t^n}^{t^{n+1}} \bE_h(t) \, dt.
\end{equation}
Computation is $\langle \bE_h \rangle_n^{n+1}$ is rendered tractable by the fact that PSATD in fact gives an expression for $\bE_h(t)$ on the \textit{entire} time interval $[t^n, t^{n+1}]$.  One can thus integrate this expression to find a formula for the Fourier transform of $\langle \bE_h \rangle_n^{n+1}$ \cite{ricketson2025explicit, shapoval2021overcoming}
\begin{equation} \label{eq:PSATD_meanfield}
\begin{split}
    \mathcal{F} \left[ \left\langle \bE_h \right\rangle_n^{n+1} \right] &= \frac{S}{kc \Dt} \mathcal{F} \left[ \bE^n \right] + i \frac{1-C}{k\Dt} \widehat{\mathbf{k}} \times \mathcal{F} \left[ \bB^n \right] - \frac{1 - C}{k^2 c^2 \Dt} \mathcal{F} \left[ \bj^{*, n+1/2} \right] \\
    &\qquad + \left( 1 - \frac{S}{kc\Dt} \right) \widehat{\mathbf{k}} \left( \widehat{\mathbf{k}} \cdot \mathcal{F} \left[ \bE^n \right] \right) \\
    &\qquad + \widehat{\mathbf{k}} \left( \widehat{\mathbf{k}} \cdot \mathcal{F} \left[ \bj^{*,n+1/2} \right] \right) \left( \frac{1-C}{k^2 c^2 \Dt } - \frac{\Dt}{2} \right).
\end{split}
\end{equation}
With these definitions in hand, the conclusion of \eqref{eq:field_id} still holds, and is obtained through similar logic, so exact energy conservation is again achieved.  

In addition to energy conservation, the scheme is second-order accurate by the following logic.  Define
\begin{equation} \label{eq:classical_gamma}
    \Gamma_p^n = \sqrt{ 1 + 2\frac{ \left( \bv_p^* - \frac{\bv_p^\dagger + \bv_p^n}{2} \right) \cdot (\bv_p^\dagger - \bv_p^n) }{\| \bv_p^\dagger \|^2}}
\end{equation}
so that $\bv_p^{n+1} = \Gamma_p^n \bv_p^\dagger$.  $\bv_p^\dagger$ is obtained using an explicit midpoint scheme that is trivially second-order accurate, so the scheme remains second-order if $\bv_p^{n+1}$ differs from $\bv_p^\dagger$ by at most $\order{\Dt^3}$ (an extra order is required since we're dealing with local truncation error).  It can be shown that, for the versions of the scheme described here, $\Gamma_p^n = 1 + \order{\Dt^4}$, which trivially implies $\| \bv_p^{n+1} - \bv_p^\dagger \| = \order{\Dt^4}$.  

A minor caveat in this scheme is that there is no guarantee that $\Gamma_p^n$ is real, meaning it can give rise to rare particles with unphysical, imaginary velocities.  The fraction of such particles is small simply because of the fact that $\Gamma_p^n$ is asymptotically close to unity.  This fraction was further quantified in \cite{ricketson2025explicit}, where is was shown that it scales like $\order{\Dt^{2d}}$, with $d$ the velocity-space dimension of the problem.  That analysis will apply equally well to the relativistic scheme proposed here.

In practice, for the few particles with imaginary values of $\Gamma_p^n$, we artificially set $\Gamma_p^n = 1$ as this still results in second-order temporal accuracy.  This has been observed to still admit fractional energy errors of $10^{-10}$ or better in benchmark problems with reasonable time-step sizes.  


\section{The method}
\label{sec:method}


As in the non-relativistic case, when incorporating relativistic effects it is simplest and most natural to begin with a Crank-Nicolson discretization of Maxwell's equations.  We thus propose the following relativistic generalization of the scheme \eqref{eq:EMbasic}:
\begin{equation} \label{eq:EMbasic_relativistic}
\begin{split}
    \bx_p^* &= \bx_p^n + \frac{\Delta t}{2} \frac{\bu_p^n}{\gamma_p^n}, \\
    \bE_h^* &= \bE_h^n + \frac{\Delta t}{2} \left( c^2 \nabla_h \times \bB_h^n -  \bj_h^{n,*} \right), \\
    \bB_h^* &= \bB_h^n - \frac{\Delta t}{2} \nabla_h \times \bE_h^n, \\
    \bu_p^* &= \bu_p^n + \frac{\Delta t}{2} \left( \bE_p^{*,*} + \frac{\bu_p^*}{\gamma_p^*} \times \bB_p^{*,*} \right), \\
    \bx_p^{n+1} &= \bx_p^n + \Delta t \frac{\bu_p^*}{\gamma_p^*}, \\
	\bE_h^{n+1} &= \bE_h^n + \Delta t \left( c^2 \nabla_h \times \bB_h^{n+1/2} -  \bj_h^{*,n+1/2} \right), \\
    \bB_h^{n+1} &= \bB_h^n - \Delta t \nabla_h \times \bE_h^{n+1/2}, \\
    \bu_p^\dagger &= \bu_p^n + \Delta t \left( \bE_p^{n+1/2} + \frac{\bu_p^*}{\gamma_p^*} \times \bB_p^{n+1/2} \right), \\
    \bu^{n+1}_p &= G^r(\bu_p^n, \bu_p^*, \bu_p^\dagger),
\end{split}
\end{equation}
All definitions from \eqref{eq:jEB_defs} carry over unchanged, with only the additional specification that $\bv_p^n = \bu_p^n / \gamma_p^n$ and $\bv_p^* = \bu_p^*/\gamma_p^*$, where $\gamma_p^* = \sqrt{ 1 + \left\| \bu_p^* \right\|^2/c^2}$ and similar for $\gamma_p^n$.  The function $G^r$ is momentarily unspecified, but will be the central object of the development below.  

Maxwell's equations are unmodified when relativistic effects are taken into account, so it is unsurprising that the field update steps are identical in \eqref{eq:EMbasic_relativistic} and \eqref{eq:EMbasic}.  We thus focus here on the particle push, knowing that if we can relate the one-step change in kinetic energy to $\sum_h \bE_h^{n+1/2} \cdot \bj_h^{*,n+1/2}$, the field-related portions of the energy conservation derivation -- see \eqref{eq:field_id} -- will carry through without modification.  In addition, we work with $\bE_p^{n+1/2}$, secure in the knowledge that conservation can be recovered with PSATD by replacing it with $\langle \bE_p \rangle^{n+1}_n$ as in Section \ref{sec:classical_scheme}.  We will give specifications of the scheme in FDTD and PSATD forms at the end of this section.  

Note that the update equation for $\bu^*_p$ is nonlinearly implicit, but in exactly the same sense as the Higuera-Cary scheme, this being the natural translation of that scheme to our context.  The nonlinear system is thus analytically solvable in exactly the same manner.  We outline that procedure in Appendix A for completeness.  The particle update is thus effectively explicit in the same sense as the Boris, Vay, and Higuera-Cary schemes.  

We now move to the definition of $G^r$: it is defined as the solution of the optimization problem
\begin{equation} \label{eq:opt_prob}
    G^r \left( \bu_p^n, \bu_p^*, \bu_p^\dagger \right) = \argmin_{\bu} \left\| \bu - \bu_p^\dagger \right\|^2 \quad \text{s.t.} \quad \bv_p^* \cdot \left( \bu_p^{\dagger} - \bu_p^n \right) = \left( \gamma(\bu) - \gamma_p^n \right) c^2,
\end{equation}
with $\gamma(\bu) = \sqrt{ 1 + \left\|\bu \right\|^2/c^2 }$. To see why this is a logical choice, note that
\begin{equation} \label{eq:opt_reasoning}
\begin{split}
    \Delta t | \mathbf{h} | \sum_h \mathbf{E}_h^{n+1/2} \cdot \bj_h^{*, n+1/2} &= \Delta t \sum_p w_p \bv_p^* \cdot \sum_h \bE_h^{n+1/2} S^h \left( \bx_p^{n+1/2} - \bx_h \right) \\ 
    &= \Delta t \sum_p w_p \bv_p^* \cdot \bE_p^{n+1/2} \\
    &= \sum_p w_p \bv_p^* \cdot \left( \bu_p^\dagger - \bu_p^n \right).
\end{split}
\end{equation}
Thus, the constraint in the optimization problem that defines $\bu_p^{n+1}$ enforces
\begin{equation}
\begin{split}
    \Delta t | \mathbf{h} | \sum_h \mathbf{E}_h^{n+1/2} \cdot \bj_h^{*, n+1/2} &= \sum_p w_p c^2 \left( \gamma_p^{n+1} - \gamma_p^n \right) \\
    &= \underbrace{\sum_p w_p c^2 \left( \gamma_p^{n+1} - 1 \right)}_{KE^{n+1}} - \underbrace{\sum_p w_p c^2 \left( \gamma_p^n - 1 \right)}_{KE^n},
\end{split}
\end{equation}
where $KE^n$ denotes kinetic energy at time-step $n$.  The remainder of the derivation of energy conservation, in which one shows that the grid-sum of $\bE \cdot \bj$ is related to the one-step change in potential energy, carries through completely unchanged from the non-relativistic case  -- again, because Maxwell's equations and their discretization are unmodified.  So, the constraint in \eqref{eq:opt_prob} does indeed enforce exact total energy conservation.  

As in the non-relativistic case, the objective function being minimized in \eqref{eq:opt_prob} serves both to specify a unique value of $\bu_p^{n+1}$ -- indeed, infinitely many values of $\bu$ satisfy the constraint -- and to preserve convergence in the limit of small time-step.  The value $\bu_p^\dagger$ is already a second-order accurate estimate of the proper velocity at time $t^{n+1}$, so it is sensible to ask that $\bu_p^{n+1}$ remain close to that value.  

It remains to show both that (a) the optimization problem in \eqref{eq:opt_prob} can be analytically solved so that the scheme's computational cost remains comparable to other explicit methods, and (b) the resulting value of $\bu_p^{n+1}$ is in fact second-order accurate.  As an additional verification exercise, we show that the scheme reduces to the non-relativistic one introduced in \cite{ricketson2025explicit} as $\bu / c \rightarrow 0$ (i.e.\ the non-relativistic limit).

As a brief aside before proceeding, we note that local charge conservation is also an important issue in PIC schemes.  By local charge conservation, we mean exact satisfaction of a discrete continuity equation
\begin{equation}
    \frac{\rho_h^{n+1} - \rho_h^n}{\Delta t} + \nabla_h \cdot \bj_h^{*,n+1/2} = 0,
\end{equation}
which guarantees that Gauss's law is exactly satisfied at each time.  The non-relativistic scheme was shown in \cite{ricketson2025explicit} to be trivially compatible with existing charge conservation schemes developed in \cite{chen2011energy, chen2020semi, ricketson2023pseudospectral}.  Key features of these schemes are that shape functions for current and charge deposition are different, and particle trajectories within a time-step must be decomposed cell-wise.  Because the continuity equation and current deposition are unmodified by relativistic effects, these arguments extend trivially to the relativistic case.  Since the derivations are quite cumbersome, we refer to the references above for more detail.  


\subsection{Optimization solution}

We proceed using Lagrange multipliers: the minimizer $\bu_p^{n+1}$ satisfies
\begin{equation}
    \left. \nabla_{\bu} \left\{ \frac{1}{2}\left\| \bu - \bu_p^\dagger \right\| + \lambda \bv_p^* \cdot \left( \bu_p^\dagger - \bu_p^n \right) - \lambda\left( \gamma(\bu) - \gamma_p^n \right)c^2\right\} \right\rvert_{\bu=\bu_p^{n+1}} = 0,
\end{equation}
where the factor of $1/2$ is introduced for convenience and does not move the minimizer.  Computing the derivative and doing some mild rearranging yields
\begin{equation}
    \left( 1 - \frac{\lambda}{\sqrt{1 + \left\|\bu^{n+1}\right\|^2/c^2}} \right) \bu^{n+1} = \bu^\dagger,
\end{equation}
where we have suppressed the $p$ subscript when no confusion results.  We see that the minimizer is a scalar multiple of $\bu^\dagger$.  We call that scalar $\Gamma^n$ as in \cite{ricketson2025explicit}.  Note that $\Gamma^n$ is now a function of the minimizer, but we may still assign it a name and compute it in terms of the other known velocities, since they specify $\bu^{n+1}$.  So, we proceed with substituting $\bu^{n+1} = \Gamma^n \bu^\dagger$ into the constraint in \eqref{eq:opt_prob} to find an explicit formula for $\Gamma^n$.  A few lines of straightforward algebra bring us to
\begin{equation}
    \Gamma^n = \frac{c}{\left\| \bu^\dagger \right\|} \left[ \left( \gamma^n + \frac{\bv^* \cdot \left( \bu^\dagger - \bu^n \right)}{c^2} \right)^2 - 1 \right]^{1/2}.
\end{equation}

While this is a satisfactory expression for implementation in code, it will facilitate later analysis to rearrange this expression.  Note first that
\begin{equation}
\begin{split}
    \frac{\bv^* \cdot \left( \bu^\dagger - \bu^n \right)}{c^2} &= \frac{ \left( \bv^* - \frac{\bu^\dagger + \bu^n}{\gamma^\dagger + \gamma^n} + \frac{\bu^\dagger + \bu^n}{\gamma^\dagger + \gamma^n} \right) \cdot \left( \bu^\dagger - \bu^n \right)}{c^2} \\
    &= \frac{ \left( \bv^* - \frac{\bu^\dagger + \bu^n}{\gamma^\dagger + \gamma^n} \right) \cdot \left( \bu^\dagger - \bu^n \right)}{c^2} + \frac{1}{c^2} \frac{\left\| \bu^\dagger \right\|^2 - \left\| \bu^n \right\|^2}{\gamma^\dagger + \gamma^n} \\
    &= \frac{ \left( \bv^* - \frac{\bu^\dagger + \bu^n}{\gamma^\dagger + \gamma^n} \right) \cdot \left( \bu^\dagger - \bu^n \right)}{c^2} + \frac{(\gamma^\dagger)^2 - (\gamma^n)^2}{\gamma^\dagger + \gamma^n} \\
    &= \frac{ \left( \bv^* - \frac{\bu^\dagger + \bu^n}{\gamma^\dagger + \gamma^n} \right) \cdot \left( \bu^\dagger - \bu^n \right)}{c^2} + \gamma^\dagger - \gamma^n.
\end{split}
\end{equation}
Substituting this into the expression for $\Gamma^n$ above, and also bringing the factor of $c/\| \bu^\dagger \|$ inside the square root and expressing it in terms of $\gamma^\dagger$, gives
\begin{equation}
    \Gamma^n = \left[ \frac{\left( \gamma^\dagger +  \left( \bv^* - \frac{\bu^\dagger + \bu^n}{\gamma^\dagger + \gamma^n} \right) \cdot \frac{\left( \bu^\dagger - \bu^n \right)}{c^2}\right)^2 - 1}{\left( \gamma^\dagger \right)^2 - 1} \right]^{1/2}.
\end{equation}
In the name of brevity, define 
\begin{equation} \label{eq:delta_def}
    \delta = \left( \bv^* - \frac{\bu^\dagger + \bu^n}{\gamma^\dagger + \gamma^n} \right) \cdot \frac{\left( \bu^\dagger - \bu^n \right)}{c^2}.
\end{equation}
Then, expanding the square in the expression for $\Gamma$ and canceling terms gives
\begin{equation} \label{eq:Gamma_formula}
    \Gamma^n = \left[ 1 + \frac{2\delta \gamma^\dagger + \delta^2}{\left( \gamma^\dagger \right)^2 - 1} \right]^{1/2}.
\end{equation}
This expression makes it abundantly clear that understanding the proximity of $\Gamma$ to unity comes down to understanding the size of $\delta$.  Analyzing the scaling of $\delta$ with time-step will be the crux of the next subsection in which we understand the order of the scheme.  

Before proceeding, we note that as in the non-relativistic case \cite{ricketson2025explicit}, there is no guarantee that $\Gamma$ is a real number.  $\delta$ may be negative, and in some rare cases this may lead to $\Gamma$ taking on an imaginary value. As noted above, the frequency of such occurrences was quantified in \cite{ricketson2025explicit}, and that analysis is unmodified by moving to the relativistic regime.  

\subsection{Reproducing the non-relativistic case}
Writing $\left( \gamma^\dagger \right)^2 - 1$ in terms of velocities, we can find
\begin{equation}
    \frac{\delta}{\left( \gamma^\dagger \right)^2 - 1} = \left( \bv^* - \frac{\bu^\dagger + \bu^n}{\gamma^\dagger + \gamma^n} \right) \cdot \frac{\left( \bu^\dagger - \bu^n \right)}{\left\| \bu^\dagger \right\|^2}.
\end{equation}
In the non-relativistic limit, i.e. $\bv/c \rightarrow 0$, one has $\bu \rightarrow \bv$, so we get 
\begin{equation}
    \frac{\delta}{\left( \gamma^\dagger \right)^2 - 1} \rightarrow \frac{\left( \bv^* - \frac{\bv^\dagger + \bv^n}{2} \right) \cdot \left( \bv^\dagger - \bv^n \right)}{\left\| \bv^\dagger \right\|^2}.
\end{equation}
On the other hand, 
\begin{equation}
    \frac{\delta^2}{\left( \gamma^\dagger \right)^2 - 1} = \frac{1}{\left\| \bu^\dagger \right\|^2 c^2} \left[ \left( \bv^* - \frac{\bu^\dagger + \bu^n}{\gamma^\dagger + \gamma^n} \right) \cdot \left( \bu^\dagger - \bu^n \right) \right]^2 \rightarrow 0,
\end{equation}
where this goes to zero in the non-relativistic limit because all velocities are negligible compared to $c$.  

So, we get that in the non-relativistic limit 
\begin{equation}
    \Gamma^n \rightarrow \left[ 1 + 2 \frac{\left( \bv^* - \frac{\bv^\dagger + \bv^n}{2} \right) \cdot \left( \bv^\dagger - \bv^n \right)}{\left\| \bv^\dagger \right\|^2} \right]^{1/2},
\end{equation}
which is exactly the non-relativistic expression \eqref{eq:classical_gamma}.

\subsection{Accuracy}

As already discussed, $\bu^\dagger$ is a second-order accurate approximation of the (proper) velocity at time $t^{n+1}$.  To establish that $\bu^{n+1}$ is as well, it suffices to show that $\| \bu^{n+1} - \bu^\dagger \| = \mathcal{O}\left(\Delta t^3 \right)$.  For this, it suffices to show that $\Gamma^n = 1 + \mathcal{O}\left( \Delta t^3 \right)$.  

Recall that in the non-relativistic case, we were actually able to show that $\Gamma^n = 1 + \mathcal{O} \left( \Delta t^4 \right)$ for the version of the scheme considered here.  This has the benefit of reducing the instances in which $\Gamma$ is imaginary, thus improving energy conservation in practice.  We are not able to show a result quite this strong here, but settle instead for $\Gamma^n = 1 + \order{\Dt^3}$.  The derivation appears in Appendix B. However, the $\order{\Dt^3}$ term arises exclusively from a difference of Lorentz factors, and thus is only non-negligible when velocities are large.  

That is to say, $\Gamma^n$ differs from unity by only $\mathcal{O} \left( \Delta t^4 \right)$ when velocities are non-relativistic, but for large velocities may differ from unity by $\mathcal{O} \left( \Delta t^3 \right)$.  In both cases, the scheme remains second-order accurate.  We merely desire the improved scaling to minimize the frequency of imaginary values of $\Gamma^n$.  However, as discussed at length in \cite{ricketson2025explicit}, imaginary values of $\Gamma^n$ are far more likely for \textit{extremely small} velocities, since this makes the denominator appearing in \eqref{eq:Gamma_formula} small.  Thus, we find degraded scaling in $\Delta t$ that occurs only at \textit{large} velocities to be quite tolerable.  

It is possible to achieve $\Gamma^n = 1 + \order{\Delta t^4}$ by making significant modifications to the scheme.  Namely, we show in Appendix C that if instead of using any of the well-studied definitions of $\bar{\bv}_p$ defined in \eqref{eq:relBoris_vbardef}, \eqref{eq:Vay_vbardef}, and \eqref{eq:HC_vbardef} we build the scheme from the definition 
\begin{equation}
    \bar{\bv}_p = \frac{\bu_p^n + \bu_{p}^{n+1}}{\gamma_p^n + \gamma_p^{n+1}},
\end{equation}
then we do, in fact, achieve $\Gamma^n = 1 + \order{\Delta t^4}$. Note that this choice of $\bar{\bv}_p$ still results in second-order accuracy but is, to the best of our knowledge, unstudied.  However, we see \textit{no} improvement whatsoever in energy conservation with this version of the scheme in any of the numerical tests detailed in Section \ref{sec:numerics}.  This lends further credence to our assessment above that the $\order{\Delta t^3}$ scaling above is acceptable because it only affects particles with large velocities.  We prefer the scheme described here in the main text due to its analogy to the Higuera-Cary integrator, which has well-known structure preserving properties, but report the alternative scheme's derivation in Appendix C as a point of academic interest.  

\subsection{FDTD and PSATD versions}
The scheme \eqref{eq:EMbasic_relativistic} uses the Crank-Nicolson discretization of Maxwell's equations.  As in the non-relativistic case, the developments here can be straightforwardly applied to FDTD and PSATD Maxwell discretizations, with the latter only requiring modification of the evaluation of $\bE$ as described in Section \ref{sec:classical_scheme}.  We record these versions of the scheme here for completeness, with the derivations of energy conservation requiring no additional insights.  

The FDTD version is written as follows.
\begin{equation} \label{eq:EMleapfrog}
\begin{split}
    \bx_p^* &= \bx_p^n + \frac{\Delta t}{2} \frac{\bu_p^n}{\gamma_p^n}, \\
    \bB_h^{n+1/2} &= \bB_h^{n-1/2} - \Delta t \nabla_h \times \bE_h^{n}, \\
    \bE_h^* &= \bE_h^n + \frac{\Delta t}{2} \left( c^2 \nabla_h \times \bB_h^{n+1/2} -  \bj_h^{n,*} \right), \\
    \bu_p^* &= \bu_p^n + \frac{\Delta t}{2} \left( \bE_p^{*,*} + \frac{\bu_p^*}{\gamma_p^*} \times \bB_p^{*,*} \right), \\
	\bE_h^{n+1} &= \bE_h^n + \Delta t \left( c^2 \nabla_h \times \bB_h^{n+1/2} -  \bj_h^{*,n+1/2} \right), \\
    \bx_p^{n+1} &= \bx_p^n + \Delta t \frac{\bu_p^*}{\gamma_p^*}, \\
    \bu_p^\dagger &= \bu_p^n + \Delta t \left( \bE_p^{n+1/2} + \frac{\bu_p^*}{\gamma_p^*} \times \bB_p^{n+1/2} \right), \\
    \bu^{n+1}_p &= \bu_p^\dagger \sqrt{ 1 + \frac{2\delta_p \gamma^\dagger_p + \delta_p^2}{\left( \gamma_p^\dagger \right)^2 - 1}},
\end{split}
\end{equation}
with $\delta$ as defined in \eqref{eq:delta_def}.  Note that the non-standard definition of magnetic potential energy in \eqref{eq:nonstd_energy} is still required here, and for precisely the same reasons.  

The PSATD version of the scheme is written as follows.
\begin{equation} \label{eq:EM_PSATD_relativistic}
\begin{split}
    \bx_p^* &= \bx_p^n + \frac{\Delta t}{2} \frac{\bu_p^n}{\gamma_p^n}, \\
    \left( \bE_h^*, \bB_h^* \right) &= \text{PSATD}\left( \bE_h^n, \bB_h^n, \bj_h^{n,*}, \Dt/2 \right), \\
    \bu_p^* &= \bu_p^n + \frac{\Delta t}{2} \left( \bE_p^{*,*} + \frac{\bu_p^*}{\gamma_p^*} \times \bB_p^{*,*} \right), \\
    \bx_p^{n+1} &= \bx_p^n + \Delta t \frac{\bu_p^*}{\gamma_p^*}, \\
	\left( \bE_h^{n+1}, \bB_h^{n+1} \right) &= \text{PSATD} \left( \bE_h^n, \bB_h^n, \bj_h^{*, n+1/2}, \Dt \right) \\
    \bu_p^\dagger &= \bu_p^n + \Delta t \left( \left\langle \bE_p \right\rangle^{n+1}_n + \frac{\bu_p^*}{\gamma_p^*} \times \bB_p^{n+1/2} \right), \\
    \bu^{n+1}_p &= \bu_p^\dagger \sqrt{ 1 + \frac{2\delta_p \gamma^\dagger_p + \delta_p^2}{\left( \gamma_p^\dagger \right)^2 - 1}},
\end{split}
\end{equation}
where $\text{PSATD}(\bE, \bB, \bj, \Delta t)$ denotes that PSATD advancement of $\bE$ and $\bB$ by the time-step $\Delta t$ using the fixed current $\bj$ according to \eqref{eq:PSATD_update}, and $\left\langle \bE_p \right\rangle^{n+1}_n = \sum_{h} \left\langle \bE_h \right\rangle^{n+1}_n S^h(\bx_h - \bx_p^{n+1/2})$ and $\left\langle \bE_h \right\rangle^{n+1}_n$ is defined by its Fourier transform given in \eqref{eq:PSATD_meanfield}.

Note that both of these versions of the scheme are fully explicit.  

\section{Numerical results}
\label{sec:numerics}

We report results from an implementation of the scheme above on a periodic box in two spatial and two velocity dimensions.  We denote the dimensions of the box by $L_x$ and $L_y$.  The number of cells in each direction is $N_x$ and $N_y$, respectively.  Except where explicitly noted, we work in the non-dimensionlization in which length is scaled by $c/\omega_p$, so that the speed of light is normalized to unity.  

Throughout, we compare against a ``standard PIC" discretization, either with PSATD or Crank-Nicolson discretization.  By ``standard", we will mean the following scheme, written in the CN case for simplicity but readily generalizable to PSATD as in the discussion above:
\begin{equation}
\begin{split}
    \bx_p^* &= \bx_p^n + \frac{\Delta t}{2} \frac{\bu_p^n}{\gamma_p^n}, \\
    \bu_p^* &= \bu_p^n + \frac{\Delta t}{2} \left( \bE_p^{n,*} + \frac{\bu_p^*}{\gamma_p^*} \times \bB_p^{n,*} \right), \\
    \bx_p^{n+1} &= \bx_p^n + \Delta t \frac{\bu_p^*}{\gamma_p^*}, \\
	\bE_h^{n+1} &= \bE_h^n + \Delta t \left( c^2 \nabla_h \times \bB_h^{n+1/2} -  \bj_h^{*,n+1/2} \right), \\
    \bB_h^{n+1} &= \bB_h^n - \Delta t \nabla_h \times \bE_h^{n+1/2}, \\
    \bu_p^{n+1} &= \bu_p^n + \Delta t \left( \bE_p^{n+1/2} + \frac{\bu_p^*}{\gamma_p^*} \times \bB_p^{n+1/2} \right), \\
\end{split}
\end{equation}
Note that this resembles the energy conserving scheme closely, but without the correction to enforce energy conservation and without the initial half-step, ``predictor" stage for the electromagnetic fields (recall that this is what distinguishes ``version 1" and ``version 2" in \cite{ricketson2025explicit}).  The scheme is second-order accurate and still based on the Higuera-Cary particle update, thus isolating the differences between this and the new scheme to the novel pieces introduced above.  

\subsection{Relativistic two-stream instability}
\label{sec:twostream}

Our implementation is verified using a standard two-stream instability test problem.  We work in one configuration ($N_y = 1$) and two velocity dimensions. We initialize $f$ with two counter-streaming beams as follows:
\begin{equation}
\begin{split}
    f_0(x,u_x,u_y) &= \frac{1}{4\pi u_{th}^2} e^{-u_y^2/2 u_{th}} \left( e^{-(u_x - u_b)^2/2u_{th}} + e^{-(u_x + u_b)^2/2u_{th}} \right)  \\
    &\qquad \times\left( 1 + \alpha_x \cos \left( \frac{2 \pi x}{L} \right) \right).
\end{split}
\end{equation}
In the cold beam limit ($u_{th} \rightarrow 0$), the linear dispersion relation for this problem is \cite{chen2020semi} 
\begin{equation}
    \frac{\gamma_b^{-3}}{(\omega + k v_b)^2} + \frac{\gamma_b^{-3}}{(\omega - k v_b)^2} = 2,
\end{equation}
where $\gamma_b = \sqrt{ 1 + u_b^2/c^2 }$ and $v_b = u_b/\gamma_b$.  Note that the only distinction between this and the non-relativistic dispersion relation is the factor of $\gamma_b^{-3}$, which serves to reduce the growth rate at relativistic beam velocities.  

For our test, we choose $L = 2\pi$, $\alpha_x = 0.01$, $u_{th} = 0.05$, and test four values of $v_b = 0.3, 0.4, 0.5, 0.6$.  With $u_b \gg u_{th}$, we find that the cold-beam approximation accurately predicts growth rates.  

\begin{figure}
    \centering
    \includegraphics[width=1.1\linewidth]{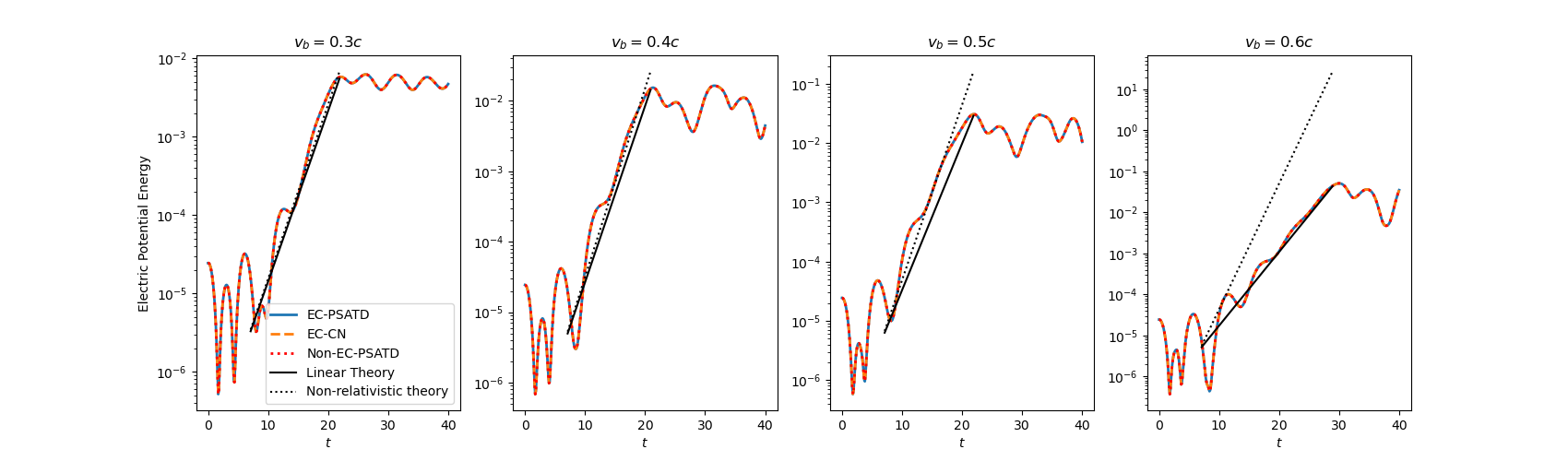}
    \caption{Electrostatic potential energy for the two-stream test cases, showing good agreement with theoretical growth rate, even when the relativistic growth rate differs markedly from the classical one.  }
    \label{fig:ts_potential}
\end{figure}

We use the discretization parameters $\Dt = 0.1$, $N_x = 32$, $1000$ particles per cell (for a total of $32,000$ particles), and run to final time $T=40$.  We utilize a quiet start in configuration space, while velocities are randomly sampled from the appropriate normal distributions.  The same random sampling is used for each scheme tested, so initial conditions are in fact identical.  In Figure \ref{fig:ts_potential}, we observe excellent agreement with the relativistic growth rate as well as with the classical limit for smaller values of $v_b$.  In addition, the energy-conserving and standard PIC schemes agree well.  

\begin{figure}
    \centering
    \includegraphics[width=0.48\linewidth]{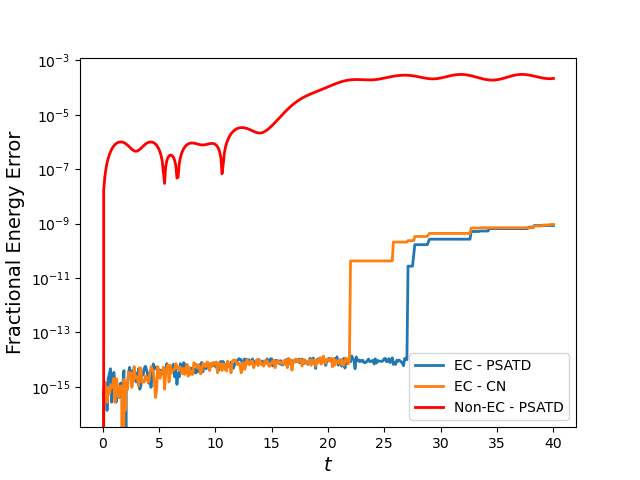}
    \includegraphics[width=0.48\linewidth]{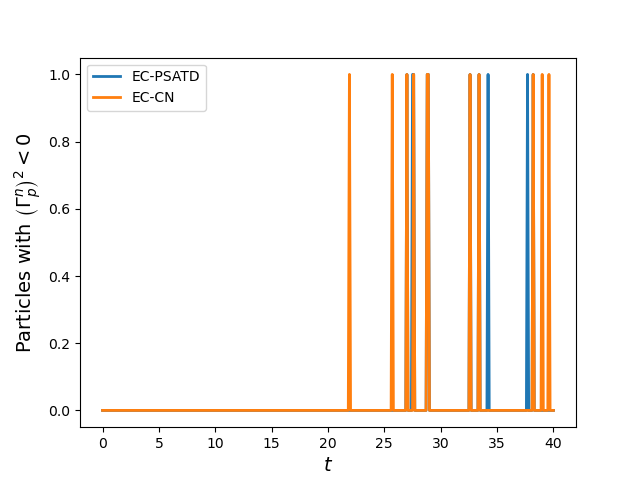}
    \caption{\underline{Left}: Fractional error in total energy as a function of time for the two-stream test problem.  Improved energy accuracy of the new scheme is readily observed. \underline{Right}: Number of particles at each time-step with imaginary $\Gamma$ values at each time-step.}
    \label{fig:ts_totalenergy}
\end{figure}

In Figure \ref{fig:ts_totalenergy} we report fractional energy errors and, for the energy-conserving schemes, the number of particles with $\Gamma_p^2 < 0$ at each time-step.  The improved energy conservation of the new scheme is readily apparent, as is the rarity of particles with imaginary $\Gamma$.  The displayed plots are for the $v_b = 0.6$ case, but other cases are not meaningfully different.  

\subsection{Relativistic Landau Damping}
\label{sec:landau}
We reproduce a test case studied in \cite{arrighi2024new}, in which a method for machine-precision evaluation of the plasma dispersion function is presented that applies to relativistic plasmas.  This results in predictions for relativistic modifications of the classical Landau damping rates.  

In contrast to other tests, we work here in a non-dimensionalization consistent with that used in \cite{arrighi2024new}.  Namely, while time is still scaled by the plasma frequency, length is scaled by Debye length and velocity by the thermal velocity.  The speed of light presented is then given in multiples of the thermal velocity, with smaller values of $c$ corresponding to more relativistic cases.  

We use $L = 6\pi$, to admit perturbations with wave-number $k=1/3$ to mirror the published damping rates of \cite{arrighi2024new}.  The initial distribution is 
\begin{equation}
    f_0(x,u_x, u_y) = \frac{1}{2\pi} e^{-(u_x^2 + u_y^2)/2} \left( 1 + \alpha \cos k x \right).
\end{equation}
We choose $\alpha = 0.05$.  Discretization parameters are $\Dt = 0.025$ (to resolve the faster light speed compared to other test problems), $N_x = 64$, final time $T = 30$, and $8000$ particles per cell to mitigate sampling noise that can affect the observation of the very small damping rates in these problems.  

\begin{figure}[h]
    \centering
    \includegraphics[width=0.75\linewidth]{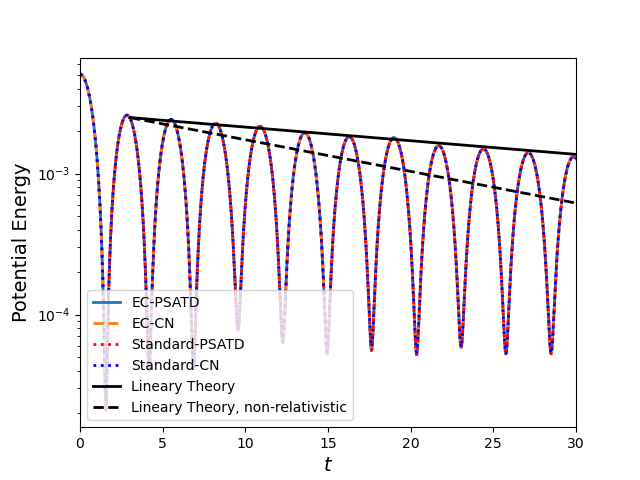}
    \caption{Potential energy as a function of time for relativistic Landau damping test case.  All tested schemes match the relativistic damping rate predicted by linear theory.}
    \label{fig:Landau_potential}
\end{figure}

Again following \cite{arrighi2024new}, we set $c=8$ -- the ``strongly relativistic" case studied there.  The predicted damping rate is $\gamma_{rel} = 0.01113948$  for the proper-velocity Maxwellian we use as our initial condition.  Meanwhile, the non-relativistic prediction of damping rate is $\gamma_{class} = 0.02587$.  Potential energy is reported in Figure \ref{fig:Landau_potential}, which shows a damping rate that matches the relativistic prediction for all schemes tested.  


\begin{figure}
    \centering
    \includegraphics[width=0.49\linewidth]{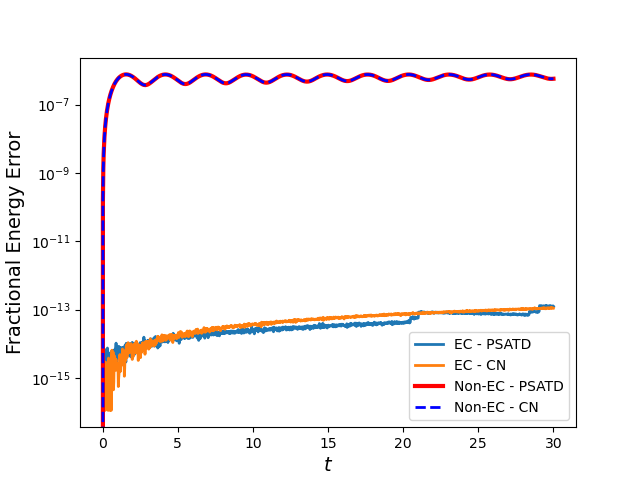}
    \includegraphics[width=0.49\linewidth]{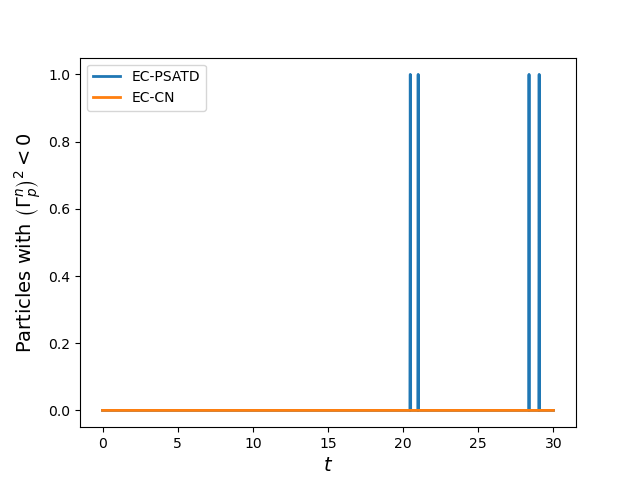}
    \caption{Energy conservation (left) and number of problematic particles at each time-step (right) for the Landau damping test case.  The expected roundoff-level energy accuracy is observed for the energy conserving schemes.}
    \label{fig:Landau_conservation}
\end{figure}

Energy conservation and the number of problematic particles for the energy-conserving schemes are reported in Figure \ref{fig:Landau_conservation}.  While standard PIC conserves energy quite well for this simple problem, 6 orders of magnitude improvement in conservation is still observed for the new energy-conserving schemes.  Problematic particles are quite rare in this example.

\subsection{Relativistic Weibel instability}
\label{sec:weibel}
We study a slightly modified version of the 1D2V Weibel instability test case of \cite{cheng2014energy}, with the modifications coming only from the inclusion of relativistic effects.  Using our nondimensionalization in which $c=1$ corresponds to the scaling used in \cite{cheng2014energy}, and we initialize the distribution function and fields according to 
\begin{equation}
\begin{split}
    f(y, u_x, u_y, t=0) &= \frac{1}{\pi \beta} e^{-u_y^2/\beta} \left[ \delta e^{-(u_x - u_{0,1})^2/\beta} + (1 - \delta) e^{-(u_x + u_{0,2})^2/\beta} \right], \\
    E_x(y, t=0) &= E_y(y, t=0) = 0, \\
    B_z(y, t=0) &= b \sin (k_0 y ).
\end{split}
\end{equation}
We mimic the parameters of Run 1 from \cite{cheng2014energy}, but with increased beam velocities $u_{0,1}$ and $u_{0,2}$ to emphasize relativistic effects. We choose 
\begin{equation}
\begin{split}
    &\beta = 0.01, \qquad \delta = 0.5, \qquad u_{0,1} = u_{0,2} = 1.25, \\
    &b = 0.001, \qquad k_0 = 0.2.
\end{split}
\end{equation}
This choice of proper velocity for the two counter-streaming beams corresponds to a physical velocity of $0.78c$, making relativistic effects quite significant.

Because the initial perturbation amplitude (controlled by $b$) is so small, and the Weibel instability saturates at relatively low amplitude, we again employ a for this test to observe the linear growth of the instability over several orders of magnitude.  In particular, the initial particle positions are specified deterministically on a uniform grid.  Particle velocities are still randomly sampled from the specified bi-Maxwellian distribution.

\begin{figure}
    \centering
    \includegraphics[width=0.75\linewidth]{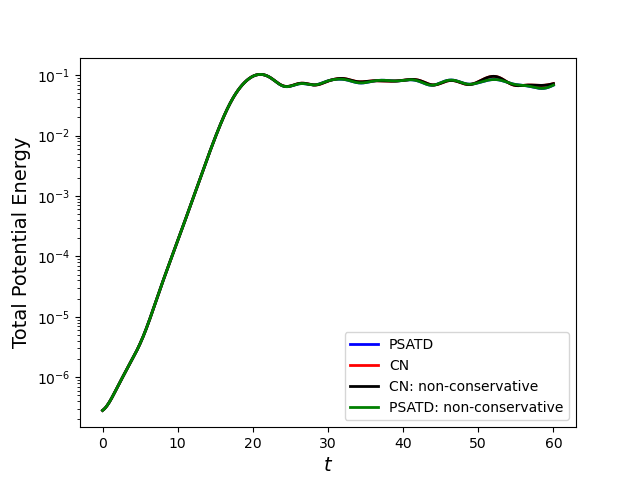}
    \caption{Total potential energy for Weibel test problem, using Crank-Nicolson and PSATD discretizations of Maxwell's equations.  We show both the new conservative schemes and classical PIC methods.  All results agree well, in that all four curves overlap.  }
    \label{fig:weibel_potential}
\end{figure}

We use $N_y = 32$, $\Dt = 0.1$, and $3200$ particles per cell.  The total potential energy for both conservative and non-conservative schemes with Crank-Nicolson and PSATD appears in Figure \ref{fig:weibel_potential}.  All schemes agree well.

\begin{figure}
    \centering
    \includegraphics[width=0.75\linewidth]{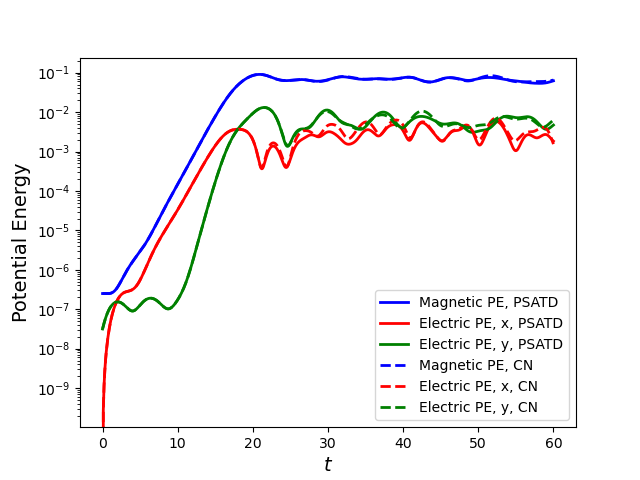}
    \caption{Breakdown of sources of potential energy in Weibel test problem.  We show only the conservative schemes here, as non-conservative schemes agree well on these axes.  }
    \label{fig:weibel_pebreakdown}
\end{figure}

\begin{figure}
    \centering
    \includegraphics[width=0.75\linewidth]{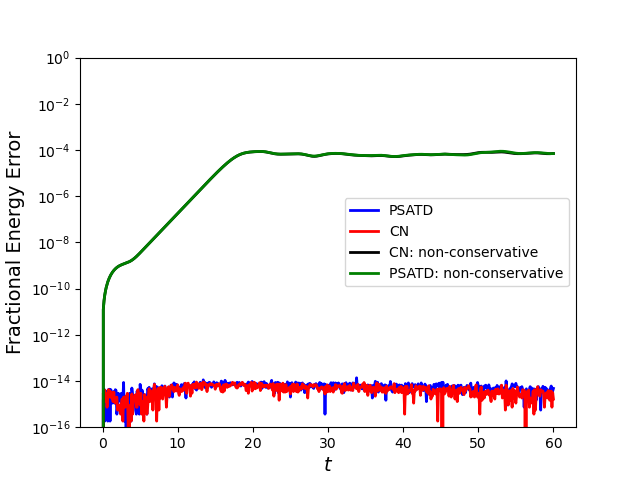}
    \caption{Fractional error in total energy for Weibel instability problem.  Improved energy accuracy of the proposed scheme is readily observed.}
    \label{fig:w_totalenergy}
\end{figure}

Some minor differences are visible in the nonlinear phase of the instability when the potential energy is broken into components from magnetic and electric potentials, similar to what is shown in \cite{cheng2014energy}.  This is shown in Figure \ref{fig:weibel_pebreakdown}.  Distinctions between the conservative and non-conservative schemes are only visible for this problem by plotting total energy conservation errors, which we do in Figure \ref{fig:w_totalenergy}.  As indicated by the fact that the conservative schemes conserve energy to double precision, we observe \textit{zero} problematic particles for this test problem, both with PSATD and Crank-Nicolson.

\subsection{Filamentation instability}
The filamentation instability presents a more challenging verification exercise because (a) it is inherently two dimensional, since the beam propagation direction and unstable wave-vectors are orthogonal, and (b) the fastest growing wave numbers occur as $k \rightarrow \infty$.  In an effort to resolve large $k$ values where the asymptotic growth rate is known, we use a small domain size $L_x = L_y = \pi$.  We initialize particles by sampling from the distribution
\begin{equation}
    f_0(u_x,u_y) = \frac{1}{4\pi} e^{-u_y^2/2 u_{th}} \left( e^{-(u_x - u_b)^2/2u_{th}} + e^{-(u_x + u_b)^2/2u_{th}} \right).
\end{equation}
Note that unlike the two-stream case, we initialize an exactly homogeneous distribution in configuration space, relying on sampling noise to trigger the instability at arbitrary wave numbers.  

With beams propagating along the $x$-axis, the instability generates density perturbations along the $y$-axis.  Again motivated by the desire to capture large wave-numbers in $y$ while keeping computational cost manageable, we choose $N_x = 32$ and $N_y = 2048$.  We use $200$ particles per cell, $\Dt = 0.1$, and show results with $u_b = 0.75$.  

We verify against the theoretical linear growth rate in the cold beam and infinite wave-number limits, given in our normalization by \cite{bret2010multidimensional}
\begin{equation}
    \delta = \frac{v_b}{\sqrt{\gamma_b}},
\end{equation}
where $\gamma_b = \sqrt{ 1 + u_b^2}$ and $v_b = u_b/\gamma_b$.  This expression differs from that in \cite{bret2010multidimensional} by a factor of $\sqrt{2}$ only because we normalize time by the plasma frequency corresponding to the overall plasma density, while they use the density of an individual beam.  

\begin{figure}
    \centering
    \includegraphics[width=0.75\linewidth]{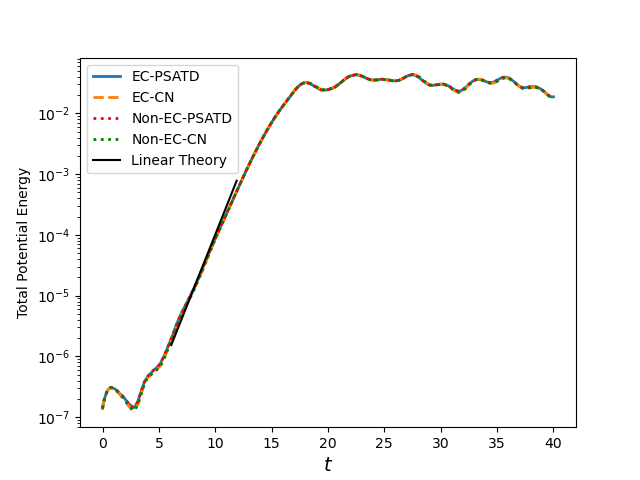}
    \caption{Growth of potential energy in the filamentation test case with cold initial beams.  Theoretical growth rate in $k \rightarrow \infty$ limit is approximately reproduced.  }
    \label{fig:PE_fil_ver}
\end{figure}

For verification, we present a case with perfectly cold beams -- i.e. $u_{th} = 0$.  The growth in magnetic potential compared to the analytically predicted growth rate is shown in Figure \ref{fig:PE_fil_ver}.
Of course, any numerical simulation can resolve only finite wave-numbers, so we are satisfied in observing that the growth rate is near, but slightly smaller than, the theoretical prediction as $k \rightarrow \infty$.  

\begin{figure}
    \centering
    \includegraphics[width=0.75\linewidth]{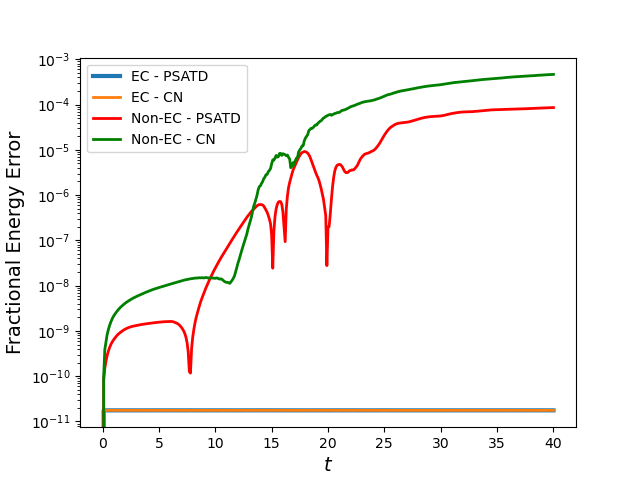}
    \caption{Fractional energy errors over time for the filamentation test case with cold initial beams.  Excellent conservation is observed for the new schemes with both spatial discretizations.}
    \label{fig:filamentation_econs}
\end{figure}

We also plot energy conservation for each scheme in Figure \ref{fig:filamentation_econs}.  As in the Weibel test case, no problematic particles are observed in the entire simulation run, and the new scheme thus features excellent conservation, improving on standard schemes by as much as 7 orders of magnitude.  

\begin{figure}[h]
    \centering
    \includegraphics[width=1.1\linewidth]{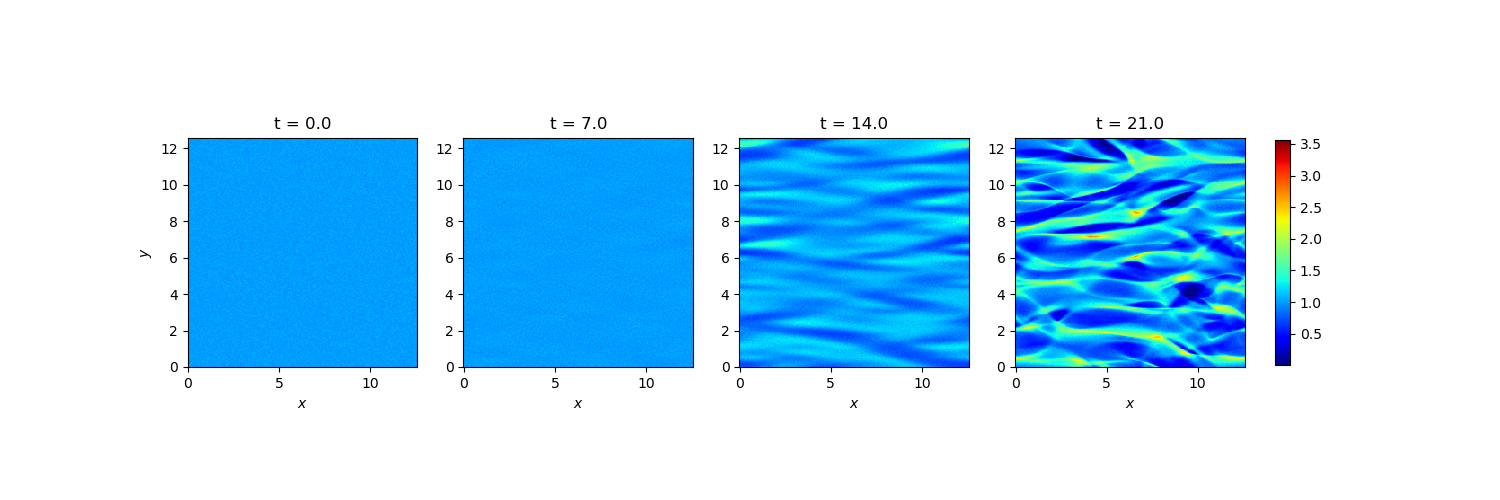}
    \caption{Temporal snapshots of electron density in a filamentation instability test case with $u_{th} = 0.05$, showing formation and breakup of the eponymous ``filaments".}
    \label{fig:filamentation_snapshots}
\end{figure}

\begin{figure}
    \centering
    \includegraphics[width=0.75\linewidth]{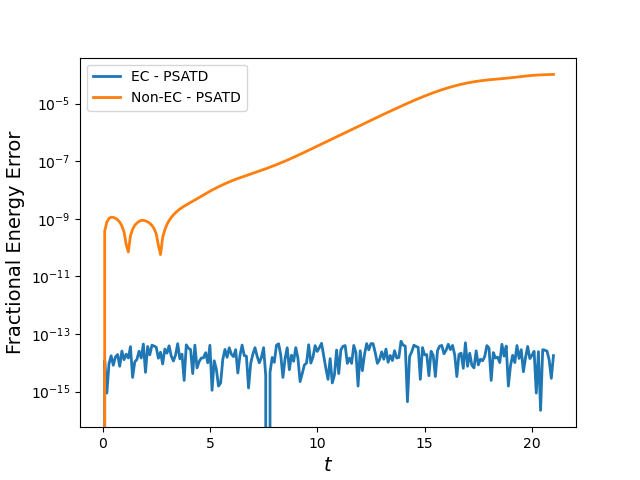}
    \caption{Energy error as a function of time for the filamentation instability test case with $u_{th} = 0.05$.}
    \label{fig:filamentation_bigConservation}
\end{figure}

A more interesting test case uses non-zero $u_{th} = 0.05$ and a larger domain $L_x = L_y = 4\pi$, for break-up of the ``filaments" formed by the instability.  In Figure \ref{fig:filamentation_snapshots} we show density snapshots from a run with $N_x = N_y = 256$ and $800$ particles per cell that illustrate the formation of filaments, their finite width induced by thermal effects, and their breakup in the nonlinear phase of the instability.  Energy conservation for this test case with PSATD -- both with standard PIC and the new energy-conserving method -- are reported in Figure \ref{fig:filamentation_bigConservation}, again showing roundoff-level energy accuracy achieved by the new scheme.

\section{Conclusions}
\label{sec:conclusions}

We have extended our earlier explicit, energy conserving PIC scheme \cite{ricketson2025explicit} to apply to relativistic plasmas.  We have shown that, as in the classical case, the analytic solution of a local optimization problem for each particle may be used to enforce exact energy conservation.  The formulation of that optimization problem and its solution are described, as is the recovery of the non-relativistic limit.  As before, the optimization is not guaranteed to admit a real solution, but we show that such issues are sufficiently rare to admit round-off level energy accuracy in many practical simulations.  As in the non-relativistic case, we show that the scheme is compatible with widely used spatial discretizations for Maxwell's equations.  

Opportunities for future work are myriad, and include application to more challenging problems as well as extension to relativistic collisional plasmas, building on \cite{yoo2025explicit}.  

\appendix


\section*{Acknowledgements}
The authors wish to acknowledge valuable private communication with Luis Chac\'{o}n and Andrew Christlieb.  This work was performed under the auspices of the U.S. Department of Energy by LLNL under contract DE-AC52-07NA27344. Both authors were supported by the DOE Office of Applied Scientific Computing Research (ASCR) Mathematical Multifaceted Integrated
Capabilities Center (MMICC) Program under grant DE-SC0023164.  Additionally, the work of J. Hu was partially supported by AFOSR grant FA9550-21-1-0358.

\section*{Appendix A}
\label{sec:App_VHCtrick}
It is instructive to first rederive the fact that, if any three vectors $\bv, \mathbf{a}, \mathbf{b} \in \mathbb{R}^3$ are related by
\begin{equation} \label{eq:vac_init}
    \bv = \mathbf{a} + \bv \times \bb,
\end{equation}
one can solve for $\bv$ explicitly and find
\begin{equation} \label{eq:vac_final}
    \bv = \frac{ \mathbf{a} + (\mathbf{a} \cdot \bb) \bb + \mathbf{a} \times \bb}{1 + b^2}.
\end{equation}
To see why, decompose $\bv$ and $\mathbf{a}$ into components parallel and perpendicular to $\mathbf{b}$: $\bv = \bv_\parallel + \bv_\perp$ and similar for $\mathbf{a}$.  Trivially, $\bv_\parallel = \mathbf{a}_\parallel = \mathbf{b} ( \mathbf{a} \cdot \mathbf{b} ) / b^2$.  Crossing the perpendicular component of \eqref{eq:vac_init} with $\mathbf{b}$ and noting that $(\bv_\perp \times \mathbf{b}) \times \mathbf{b} = - b^2 \bv_\perp$, we have
\begin{equation}
    \bv \times \mathbf{b} = \bv_\perp \times \mathbf{b} = \mathbf{a} \times \mathbf{b} - \bv_\perp b^2.
\end{equation}
Substituting this back into the perpendicular component of \eqref{eq:vac_init} gives
\begin{equation}
    \bv_\perp = \mathbf{a}_\perp + \mathbf{a} \times \mathbf{b} - b^2 \bv_\perp \qquad \implies \qquad \bv_\perp = \frac{\mathbf{a}_\perp + \mathbf{a} \times \mathbf{b}}{1 + b^2}.
\end{equation}
Combining this with our observation about the parallel component, we have
\begin{equation}
    \bv = \frac{\mathbf{a} \cdot \mathbf{b}}{b^2} \mathbf{b} + \frac{ \mathbf{a} - \frac{\mathbf{a} \cdot \mathbf{b}}{b^2} \mathbf{b} + \mathbf{a} \times \mathbf{b}}{1 + b^2},
\end{equation}
which simplifies to \eqref{eq:vac_final}.  

Next, consider the proposed update for $\bu_p^*$:
\begin{equation} \label{eq:ustar_update}
    \bu_p^* = \bu_p^n + \frac{\Delta t}{2} \left( \bE_p^{*,*} + \frac{\bu_p^*}{\gamma_p^*} \times \bB_p^{*,*} \right).
\end{equation}
Letting $\bar{\mathbf{a}} = \bu_p^n + \frac{\Delta t}{2} \bE_p^{*,*}$ and $\bar{\mathbf{b}} = \frac{\Dt}{2}\bB_p^{*,*} $, \eqref{eq:vac_final} implies
\begin{equation} \label{eq:upstar_expr}
    \bu_p^* = \frac{\bar{\mathbf{a}} + (\bar{\mathbf{a}} \cdot \bar{\mathbf{b}}) \bar{\bb} / (\gamma_p^*)^2 + \bar{\mathbf{a}} \times \bar{\mathbf{b}}/\gamma_p^*}{1 + \bar{b}^2 / (\gamma_p^*)^2}.
\end{equation}

Next, note that dotting \eqref{eq:ustar_update} with $\bu_p^*$ implies that $\| \bu_p^* \|^2 = \bar{\mathbf{a}} \cdot \bu_p^*$.  Substituting in the expression above for $\bu_p^*$ on the right and noting that $\| \bu_p^* \|^2 = c^2 ((\gamma_p^*)^2 - 1)$, we have
\begin{equation}
    c^2 \left((\gamma_p^*)^2 - 1 \right) = \frac{ \bar{a}^2 + (\bar{\mathbf{a}} \cdot \bar{\bb})^2 / (\gamma_p^*)^2 }{1 + \bar{b}^2 / (\gamma_p^*)^2}.
\end{equation}
Multiplying through by factors of $\gamma_p^*$ as appropriate, we arrive at a quadratic equation for $(\gamma_p^*)^2$:
\begin{equation}
\begin{split}
    &(\gamma_p^*)^4 + \left( \bar{b}^2 - 1 - \bar{a}^2/c^2 \right) (\gamma_p^*)^2 - \bar{b}^2 - (\bar{\mathbf{a}} \cdot \bar{\bb})^2/c^2 = 0,
\end{split}
\end{equation}
whose (positive) solution is
\begin{equation}
    (\gamma_p^*)^2 = \frac{1}{2} \left( \gamma^2(\bar{\mathbf{a}}) - \bar{b}^2 + \sqrt{  \left( \gamma^2(\bar{\mathbf{a}}) - \bar{b}^2 \right)^2 + 4 \left( \bar{b}^2 + (\bar{\mathbf{a}} \cdot \bar{\bb})^2/c^2 \right)}\right),
\end{equation}
where $\gamma^2(\bar{\mathbf{a}}): = 1 + \bar{a}^2/c^2$.
The positive square root of this expression completely specifies $\gamma_p^*$ in terms of known quantities.  

Knowing $\gamma_p^*$, \eqref{eq:upstar_expr} now completely specifies $\bu_p^*$ in terms of known quantities.  

\section*{Appendix B}
\label{sec:App_Gammascaling}
As one may expect, the key to understanding the scaling of $\Gamma$ in \eqref{eq:Gamma_formula} is to understand the scaling of $\delta$, defined in \eqref{eq:delta_def}.  The second term in the dot product that defines $\delta$ is manifestly $\mathcal{O}(\Delta t)$, so we concern ourselves primarily with the first.  We begin by noticing that 
\begin{equation}
\begin{split}
    \bv^* - \frac{\bu^\dagger + \bu^n}{\gamma^\dagger + \gamma^n}  &= \frac{\bu^*}{\gamma^*} - \frac{\bu^\dagger + \bu^n}{\gamma^\dagger + \gamma^n} \\
    &= 2\frac{\frac{\gamma^\dagger + \gamma^n}{2} \bu^* - \gamma^* \frac{\bu^\dagger + \bu^n}{2}}{\gamma^* \left( \gamma^\dagger + \gamma^n \right)} \\
    &= \frac{2}{\gamma^* \left( \gamma^\dagger + \gamma^n \right)} \left[ \bu^* \left( \frac{\gamma^\dagger + \gamma^n}{2} - \gamma^* \right) + \gamma^* \left( \bu^* - \frac{\bu^\dagger + \bu^n}{2}\right) \right].
\end{split}
\end{equation}

By precisely the same logic used in \cite{ricketson2025explicit}, the difference of velocities in the last line is $\mathcal{O}\left( \Delta t^3 \right)$.  Indeed, directly substituting in the definition of the (proper) velocity update gives
\begin{equation}
    \bu^* - \frac{\bu^\dagger + \bu^n}{2} = \frac{\Delta t}{2} \left( \bE^{*,*} - \bE^{n+1/2} + \bv^* \times \left( \bB^{*,*} - \bB^{n+1/2} \right) \right).
\end{equation}
The differences of fields are specifically constructed to be $\mathcal{O}\left( \Delta t^2 \right)$ -- see \cite{ricketson2025explicit} for the full derivation.

We next analyze the difference of $\gamma$'s.  Taylor expansion to second order and diligent but straightforward algebraic manipulation tells us that 
\begin{equation} \label{eq:gamma_expansion}
    \sqrt{1 + \frac{|\bu + \mathbf{a}|^2}{c^2}} = \gamma(\bu) + \frac{\mathbf{a} \cdot \bu / c^2}{\gamma(u)} + \frac{1}{2 \gamma(u) c^2} \left( \| \mathbf{a} \|^2 - \frac{(\mathbf{a} \cdot \bv )^2}{c^2}\right)  +\mathcal{O}\left(a^3\right),
\end{equation}
for arbitrary vectors $\mathbf{a}$ and $\bu$.  Here, $\gamma(\bu)$ is simply the Lorentz factor evaluated at $\bu$, and $\bv = \bu/\gamma(\bu)$.  Applying this formula to Taylor expand both $\gamma^\dagger$ and $\gamma^n$ about $\bu^*$, we find that
\begin{equation}
\begin{split}
    \frac{\gamma^\dagger + \gamma^n}{2} - \gamma^* &= \frac{\bu^*}{c^2 \gamma^*} \cdot \left[ \frac{\bu^\dagger + \bu^n}{2} - \bu^* \right] \\
    &\quad + \frac{1}{2\gamma^* c^2} \sum_{r=n,\dagger} \left\{\left\| \bu^r - \bu^* \right\|^2 - \frac{((\bu^r - \bu^*) \cdot \bv^*)^2}{c^2}\right\} \\
    &\quad + \mathcal{O}\left( \Delta t^3 \right).
\end{split}
\end{equation}
The term in the first line is $\mathcal{O}(\Delta t^3)$, since it features the same difference of velocities that was considered above.  The second line is manifestly $\order{\Dt^2}$ and does not vanish.  Overall, we thus have 
\begin{equation}
     \bv^* - \frac{\bu^\dagger + \bu^n}{\gamma^\dagger + \gamma^n} = \order{\Dt^2},
\end{equation}
with the leading order term \textit{only} coming from a difference of Lorentz factors.  It follows immediately that $\delta = \order{\Dt^3}$ and that $\Gamma = 1 + \order{\Dt^3}$.



\section*{Appendix C}
\label{sec:alt_version}
It is somewhat disappointing that the $\Gamma = 1 + \order{\Dt^4}$ result has not carried over from the non-relativistic to the relativistic case in the scheme presented above.  The reason, as discussed further in Appendix B, comes down to the distinction between midpoint and trapezoidal evaluation of $\gamma$.  The use of Higuera-Cary's midpoint-based choice of $\bar{\gamma}_p$ to define $\bv_p^*$, which appears in the definition of current density, conflicts with the trapezoidal evaluation of $\gamma$ -- namely, at $\bu^\dagger$ and $\bu^n$ -- that is inherent in the energy conservation constraint.  

One is thus motivated to wonder whether replacing our update of $\bu_p^*$ with a different choice of the parameter $\bar{\bv}_p$ appearing in \eqref{eq:gen_integrator}, can lead to a scheme with $\Gamma = 1 + \order{\Dt^4}$.  It turns out this can be done.  To see this, consider the modified particle update
\begin{equation}
\begin{split}
    \bx_p^* &= \bx_p^n + \frac{\Delta t}{2} \frac{\bu_p^n}{\gamma_p^n}, \\
    \bu_p^{**} &= \bu_p^n + \Delta t \left( \bE_p^{*,*} + \underbrace{\left( \frac{\bu_p^{**} + \bu_p^n}{\gamma^{**} + \gamma_p^n} + \right)}_{\coloneqq \bv_p^*} \times \bB_p^{*,*} \right) , \\
    \bu_p^* &= \frac{1}{2} \left( \bu_p^n + \bu_p^{**} \right) \\
    \bx_p^{n+1} &= \bx_p^n + \Delta t \bv_p^*, \\
    \bu_p^\dagger &= \bu_p^n + \Delta t \left( \bE_p^{n+1/2} + \bv_p^* \times \bB_p^{n+1/2} \right), \\
    \bu_p^{n+1} &= G^r(\bu_p^n, \bu_p^*, \bu_p^\dagger).
\end{split}
\end{equation}
In this version, the definition of $\bv_p^*$ introduced in the second line is the same quantity used when computing the current density $\bj_h^{*,n+1/2}$.  In this way, the optimization problem \eqref{eq:opt_prob} that defines $G^r$ is unchanged, since \eqref{eq:opt_reasoning} is unmodified.

As a result, the expression for $\Gamma$ in \eqref{eq:delta_def} -- \eqref{eq:Gamma_formula} also carries through unchanged.  It only remains to carry out an analysis of the size of $\delta$.  With this definition of $\bv_p^*$, one has
\begin{equation}
\begin{split}
    \bv^* - \frac{\bu^\dagger + \bu^n}{\gamma^{\dagger} + \gamma^n} &=  \frac{\bu^{**} + \bu^n}{\gamma^{**} + \gamma^n} - \frac{\bu^\dagger + \bu^n}{\gamma^{\dagger} + \gamma^n} \\
    &= \frac{(\bu^{**} + \bu^n) - (\bu^\dagger + \bu^n)}{\gamma^{**} + \gamma^n} + (\bu^\dagger + \bu^n) \left( \frac{1}{\gamma^{**}+\gamma^n} - \frac{1}{\gamma^{\dagger}+\gamma^n} \right) \\
    &= \frac{\bu^{**} - \bu^\dagger}{\gamma^{**} + \gamma^n} + \frac{\bu^\dagger + \bu^n}{\left( \gamma^{**} + \gamma^n \right) \left( \gamma^{\dagger} + \gamma^n \right)} \left( \gamma^{\dagger} - \gamma^{**} \right).
\end{split}
\end{equation}
Trivially, $| \gamma^\dagger - \gamma^{**}| = \order{\left\| \bu^\dagger - \bu^{**} \right\|}$, since $\gamma$ is a Lipschitz function of $\bu$.  Thus, for this version of the scheme we have
\begin{equation}
   \left\| \bv^* - \frac{\bu^\dagger + \bu^n}{\gamma^{\dagger} + \gamma^n} \right\| = \order{ \left\| \bu^{**} - \bu^\dagger \right\|}.
\end{equation}
By construction, $\left\| \bu^{**} - \bu^\dagger \right\| = \order{\Dt^3}$, so we have $\delta = \order{\Dt^4}$. Indeed,
\begin{equation}
    \bu^{**} - \bu^\dagger = \Delta t \left( \bE^{*,*} - \bE^{n+1/2} + \bv^* \times \left( \bB^{*,*} - \bB^{n+1/2} \right) \right).  
\end{equation}
As in Appendix B and \cite{ricketson2025explicit}, the differences of fields are constructed specifically to be $\order{\Delta t^2}$, making the entire right side is trivially $\order{\Delta t^3}$.  This implies that $\Gamma = 1 + \order{\Dt^4}$ by following the logic in Appendix B.

 \bibliographystyle{elsarticle-num} 
 \bibliography{biblio}

@article{HL18,
	author = {E. Hairer and C. Lubich},
	journal = {BIT Numer. Math.},
	pages = {969--979},
	title = {{Energy behaviour of the Boris method for charged-particle dynamics}},
	volume = {58},
	year = {2018}}

@article{Yee66,
	author = {K. Yee},
	journal = {IEEE Trans. Antennas Propag.},
	pages = {302--307},
	title = {Numerical solution of initial boundary value problems involving {M}axwell's equations in istropic media},
	volume = {14},
	year = {1966}}

@article{chacon2016curvilinear,
  title={A curvilinear, fully implicit, conservative electromagnetic PIC algorithm in multiple dimensions},
  author={Chac{\'o}n, Luis and Chen, Guangye},
  journal={Journal of computational physics},
  volume={316},
  pages={578--597},
  year={2016},
  publisher={Elsevier}
}

@article{chen2011energy,
  title={An energy-and charge-conserving, implicit, electrostatic particle-in-cell algorithm},
  author={Chen, Guangye and Chac{\'o}n, Luis and Barnes, Daniel C},
  journal={Journal of Computational Physics},
  volume={230},
  number={18},
  pages={7018--7036},
  year={2011},
  publisher={Elsevier}
}

@article{chen2020semi,
  title={A semi-implicit, energy-and charge-conserving particle-in-cell algorithm for the relativistic Vlasov-Maxwell equations},
  author={Chen, Guangye and Chacon, Luis and Yin, Lin and Albright, Brian J and Stark, David James and Bird, Robert F},
  journal={Journal of Computational Physics},
  volume={407},
  pages={109228},
  year={2020},
  publisher={Elsevier}
}

@article{qin2013boris,
  title={Why is Boris algorithm so good?},
  author={Qin, Hong and Zhang, Shuangxi and Xiao, Jianyuan and Liu, Jian and Sun, Yajuan and Tang, William M},
  journal={Physics of Plasmas},
  volume={20},
  number={8},
  year={2013},
  publisher={AIP Publishing}
}

@inproceedings{lehe2018review,
  title={Review of spectral maxwell solvers for electromagnetic particle-in-cell: Algorithms and advantages},
  author={Leh{\'e}, Remi and Vay, Jean-Luc and others},
  booktitle={Proceedings of the 13th International Computational Accelerator Physics Conference, Key West, FL, USA},
  pages={20--24},
  year={2018}
}

@article{vay2013domain,
  title={A domain decomposition method for pseudo-spectral electromagnetic simulations of plasmas},
  author={Vay, Jean-Luc and Haber, Irving and Godfrey, Brendan B},
  journal={Journal of Computational Physics},
  volume={243},
  pages={260--268},
  year={2013},
  publisher={Elsevier}
}

@article{ricketson2023pseudospectral,
  title={A pseudospectral implicit particle-in-cell method with exact energy and charge conservation},
  author={Ricketson, Lee F and Chen, Guangye},
  journal={Computer Physics Communications},
  pages={108811},
  year={2023},
  publisher={Elsevier}
}

@article{cheng2014energy,
  title={Energy-conserving discontinuous Galerkin methods for the Vlasov--Maxwell system},
  author={Cheng, Yingda and Christlieb, Andrew J and Zhong, Xinghui},
  journal={Journal of Computational Physics},
  volume={279},
  pages={145--173},
  year={2014},
  publisher={Elsevier}
}

@article{gonoskov2024explicit,
  title={Explicit energy-conserving modification of relativistic PIC method},
  author={Gonoskov, Arkady},
  journal={Journal of Computational Physics},
  volume={502},
  pages={112820},
  year={2024},
  publisher={Elsevier}
}

@article{ji2023asymptotic,
  title={An asymptotic-preserving and energy-conserving particle-in-cell method for Vlasov--Maxwell equations},
  author={Ji, Lijie and Yang, Zhiguo and Li, Zhuoning and Wu, Dong and Jin, Shi and Xu, Zhenli},
  journal={Journal of Mathematical Physics},
  volume={64},
  number={6},
  year={2023},
  publisher={AIP Publishing}
}

@article{chen2015multi,
  title={A multi-dimensional, energy-and charge-conserving, nonlinearly implicit, electromagnetic Vlasov--Darwin particle-in-cell algorithm},
  author={Chen, Guangye and Chacon, Luis},
  journal={Computer Physics Communications},
  volume={197},
  pages={73--87},
  year={2015},
  publisher={Elsevier}
}

@article{lapenta2017exactly,
  title={Exactly energy conserving semi-implicit particle in cell formulation},
  author={Lapenta, Giovanni},
  journal={Journal of Computational Physics},
  volume={334},
  pages={349--366},
  year={2017},
  publisher={Elsevier}
}

@inproceedings{bacchini2019relativistic,
  title={The relativistic implicit particle-in-cell method},
  author={Bacchini, Fabio and Amaya, Jorge and Lapenta, Giovanni},
  booktitle={Journal of Physics: Conference Series},
  volume={1225},
  number={1},
  pages={012011},
  year={2019},
  organization={IOP Publishing}
}

@article{markidis2011energy,
  title={The energy conserving particle-in-cell method},
  author={Markidis, Stefano and Lapenta, Giovanni},
  journal={Journal of Computational Physics},
  volume={230},
  number={18},
  pages={7037--7052},
  year={2011},
  publisher={Elsevier}
}

@book{birdsall2018plasma,
  title={Plasma physics via computer simulation},
  author={Birdsall, Charles K and Langdon, A Bruce},
  year={2018},
  publisher={CRC press}
}

@article{barnes2021finite,
  title={Finite spatial-grid effects in energy-conserving particle-in-cell algorithms},
  author={Barnes, Daniel C and Chac{\'o}n, Luis},
  journal={Computer Physics Communications},
  volume={258},
  pages={107560},
  year={2021},
  publisher={Elsevier}
}

@article{cowan2013generalized,
  title={Generalized algorithm for control of numerical dispersion in explicit time-domain electromagnetic simulations},
  author={Cowan, Benjamin M and Bruhwiler, David L and Cary, John R and Cormier-Michel, Estelle and Geddes, Cameron GR},
  journal={Physical Review Special Topics—Accelerators and Beams},
  volume={16},
  number={4},
  pages={041303},
  year={2013},
  publisher={APS}
}

@article{blaclard2017pseudospectral,
  title={Pseudospectral Maxwell solvers for an accurate modeling of Doppler harmonic generation on plasma mirrors with particle-in-cell codes},
  author={Blaclard, G and Vincenti, H and Lehe, R and Vay, JL},
  journal={Physical Review E},
  volume={96},
  number={3},
  pages={033305},
  year={2017},
  publisher={APS}
}

@article{petropoulos1994phase,
  title={Phase error control for FD-TD methods of second and fourth order accuracy},
  author={Petropoulos, Peter G},
  journal={IEEE transactions on antennas and propagation},
  volume={42},
  number={6},
  pages={859--862},
  year={1994},
  publisher={IEEE}
}

@article{schneider2001dispersion,
  title={Dispersion of homogeneous and inhomogeneous waves in the Yee finite-difference time-domain grid},
  author={Schneider, John B and Kruhlak, Robert J},
  journal={IEEE transactions on microwave theory and techniques},
  volume={49},
  number={2},
  pages={280--287},
  year={2001},
  publisher={IEEE}
}

@incollection{gedney2011yee,
  title={Yee algorithm for Maxwell’s equations},
  author={Gedney, Stephen D},
  booktitle={Introduction to the Finite-Difference Time-Domain (FDTD) Method for Electromagnetics},
  pages={39--73},
  year={2011},
  publisher={Springer}
}

@article{shapoval2021overcoming,
  title={Overcoming timestep limitations in boosted-frame particle-in-cell simulations of plasma-based acceleration},
  author={Shapoval, Olga and Lehe, Remi and Th{\'e}venet, Maxence and Zoni, Edoardo and Zhao, Yinjian and Vay, Jean-Luc},
  journal={Physical Review E},
  volume={104},
  number={5},
  pages={055311},
  year={2021},
  publisher={APS}
}

@article{ricketson2025explicit,
  title={An explicit, energy-conserving particle-in-cell scheme},
  author={Ricketson, Lee F and Hu, Jingwei},
  journal={Journal of Computational Physics},
  pages={114098},
  volume={537},
  year={2025},
  publisher={Elsevier}
}

@article{higuera2017structure,
  title={Structure-preserving second-order integration of relativistic charged particle trajectories in electromagnetic fields},
  author={Higuera, Adam V and Cary, John R},
  journal={Physics of Plasmas},
  volume={24},
  number={5},
  year={2017},
  publisher={AIP Publishing}
}

@article{vay2008simulation,
  title={Simulation of beams or plasmas crossing at relativistic velocity},
  author={Vay, J-L},
  journal={Physics of Plasmas},
  volume={15},
  number={5},
  year={2008},
  publisher={AIP Publishing}
}

@inproceedings{boris1970relativistic,
  title={Relativistic plasma simulation-optimization of a hybrid code},
  author={Boris, Jay P and others},
  booktitle={Proc. Fourth Conf. Num. Sim. Plasmas},
  pages={3--67},
  year={1970}
}

@article{bret2010multidimensional,
  title={Multidimensional electron beam-plasma instabilities in the relativistic regime},
  author={Bret, Antoine and Gremillet, Laurent and Dieckmann, Mark Eric},
  journal={Physics of Plasmas},
  volume={17},
  number={12},
  year={2010},
  publisher={AIP Publishing}
}

@article{arrighi2024new,
  title={A new approach to the evaluation and solution of the relativistic kinetic dispersion relation and verification with continuum kinetic simulation},
  author={Arrighi, William J and Banks, Jeffrey W and Berger, RL and Chapman, Thomas and Odu, A Gianesini and Gorman, J},
  journal={Journal of Computational Physics},
  volume={508},
  pages={113001},
  year={2024},
  publisher={Elsevier}
}

@article{yoo2025explicit,
  title={{An explicit energy-conserving particle method for the Vlasov-Fokker-Planck equation}},
  author={Yoo, Jiyoung and Hu, Jingwei and Ricketson, Lee F},
  journal={arXiv preprint arXiv:2510.03960},
  year={2025}
}

@article{nishikawa2021pic,
  title={PIC methods in astrophysics: simulations of relativistic jets and kinetic physics in astrophysical systems},
  author={Nishikawa, Kenichi and Du{\c{t}}an, Ioana and K{\"o}hn, Christoph and Mizuno, Yosuke},
  journal={Living Reviews in Computational Astrophysics},
  volume={7},
  number={1},
  pages={1},
  year={2021},
  publisher={Springer}
}

@article{breizman2019physics,
  title={Physics of runaway electrons in tokamaks},
  author={Breizman, Boris N and Aleynikov, Pavel and Hollmann, Eric M and Lehnen, Michael},
  journal={Nuclear Fusion},
  volume={59},
  number={8},
  pages={083001},
  year={2019},
  publisher={IOP Publishing}
}

@article{atzeni2005fluid,
  title={Fluid and kinetic simulation of inertial confinement fusion plasmas},
  author={Atzeni, Stefano and Schiavi, Angelo and Califano, Francesco and Cattani, F and Cornolti, Fulvio and Del Sarto, D and Liseykina, TV and Macchi, Andrea and Pegoraro, Francesco},
  journal={Computer physics communications},
  volume={169},
  number={1-3},
  pages={153--159},
  year={2005},
  publisher={Elsevier}
}

@article{schmitz2026overview,
  title={An Overview of Relativistic Particle Pushers and their Extension to Arbitrary Order Accuracy},
  author={Schmitz, Holger},
  journal={arXiv preprint arXiv:2603.06509},
  year={2026}
}





\end{document}